\documentclass[fleqn,usenatbib,useAMS]{mnras}

\usepackage{newtxtext,newtxmath}

\usepackage[T1]{fontenc}

\DeclareRobustCommand{\VAN}[3]{#2}
\let\VANthebibliography\thebibliography
\def\thebibliography{\DeclareRobustCommand{\VAN}[3]{##3}\VANthebibliography}

\usepackage{hyperref}
\usepackage{subcaption}
\usepackage{graphicx}	
\usepackage{amsmath}	
\usepackage{tablefootnote}

\def\sistff{\ion{Si}{ii}~$\lambda$6\,355}
\def\sifnst{\ion{Si}{ii}~$\lambda$5\,972}
\def\cahk{\ion{Ca}{ii}~H\&K\,}

\def\phn{\phantom{0}}

\newcommand{\nickel}{$^{56}$Ni\,}

\title[Quantitative SNe~Ia spectral sequences]{Quantitative modelling of type Ia supernovae spectral time series: Constraining the explosion physics }

\author[M. R. Magee et al.]{
M. R. Magee$^{1}$\thanks{E-mail: mrmagee.astro@gmail.com}, L. Siebenaler$^{2}$,
K. Maguire$^{3}$, K. Ackley$^{1}$, T. Killestein$^{4}$
\\
$^{1}$Department of Physics, University of Warwick, Gibbet Hill Road, Coventry CV4 7AL, UK \\ 
$^{2}$Leiden Observatory, Leiden University, NL-2300 RA Leiden, Netherlands \\
$^{3}$School of Physics, Trinity College Dublin, The University of Dublin, Dublin 2, Ireland \\ 
$^{4}$Tuorla Observatory, Department of Physics and Astronomy, FI-
20014 University of Turku, Finland
}

\date{Accepted 2024 May 07. Received 2024 May 06; in original form 2024 March 26}

\pubyear{2024}

\begin{document}
\label{firstpage}
\pagerange{\pageref{firstpage}--\pageref{lastpage}}
\maketitle

\begin{abstract}
Multiple explosion mechanisms have been proposed to explain type Ia supernovae (SNe Ia). Empirical modelling tools have also been developed that allow for fast, customised modelling of individual SNe and direct comparisons between observations and explosion model predictions. Such tools have provided useful insights, but the subjective nature with which empirical modelling is performed makes it difficult to obtain robust constraints on the explosion physics or expand studies to large populations of objects. Machine learning accelerated tools have therefore begun to gain traction. In this paper, we present \texttt{riddler}, a framework for automated fitting of SNe Ia spectral sequences up to shortly after maximum light. We train a series of neural networks on realistic ejecta profiles predicted by the W7 and N100 explosion models to emulate full radiative transfer simulations and apply nested sampling to determine the best-fitting model parameters for multiple spectra of a given SN simultaneously. We show that \texttt{riddler} is able to accurately recover the parameters of input spectra and use it to fit observations of two well-studied SNe Ia. We also investigate the impact of different weighting schemes when performing quantitative spectral fitting and show that best-fitting models and parameters are highly dependent on the assumed weighting schemes and priors. As spectroscopic samples of SNe Ia continue to grow, automated spectral fitting tools such as \texttt{riddler} will become increasingly important to maximise the physical constraints that can be gained in a quantitative and consistent manner.
\end{abstract}

\begin{keywords}
	supernovae: general --- radiative transfer 
\end{keywords}



\section{Introduction}
\label{sect:intro}
Spectroscopic observations of type Ia supernovae (SNe~Ia) are a key diagnostic probe of the explosion physics. Modelling and analysis of these observations give insight into the composition, density profile, and other physical properties of the SN that are directly linked to the explosion physics (see e.g. \citealt{jerkstrand--17, sim--17} for overviews of some modelling techniques applied to SNe spectra). As spectroscopic datasets become increasingly large however, our ability to quickly and effectively model these data, so as to quantitatively extract the relevant physical parameters, has not been able to keep pace. 

\par

Multiple different modelling techniques are currently in use. Simple line identification codes (e.g. \textsc{synapps}, \citealt{synapps}) provide useful information about which elements are present within the ejecta by allowing for individual tweaking of line strengths for different ions (e.g. \citealt{sullivan--11, parrent--2012, hsiao--13}). At the other extreme are highly sophisticated, self-consistent radiative transfer codes designed to produce synthetic observables from realistic explosion models (e.g. \citealt{hoeflich-1995a, hoeflich--2003a, stella--98,stella--06, kasen-06b, artis, hillier--2012, vanrossum--2012, ergon--18}). These predictions can then be directly compared against observations of a SN to determine whether it is consistent with a given explosion scenario (e.g. \citealt{baron--12, dessart--14a, roepke--18, shen--21}). Both of these options provide their own advantages and disadvantages. Line identification codes are often too parameterised and simplistic to provide robust predictions beyond the presence or absence of particular lines, while highly sophisticated codes come with huge computational expense that makes them ill-suited for exploration of large parameter spaces.

\par

Radiative transfer codes intermediate to these two extremes have also been developed (e.g. \citealt{mazzali--lucy--93, tardis}). These codes are designed for significantly faster, empirical modelling of SNe spectra, and therefore make a few simplifying assumptions, but contain sufficient physics such that they may be used to produce a self-consistent model for interrogating the physical conditions of the ejecta. With these codes various input parameters, such as the temperature, density, and composition of the ejecta, can be used to generate synthetic spectra to directly confront the observations. The speed and flexibility of such codes means that they can be used to fit samples of SNe~Ia \citep{mazzali--07}. Investigating how changes to these parameters affect the quality of the fit can also provide useful insights (e.g. \citealt{stehle--05, mazzali--08, tanaka--11, mazzali--14, sasdelli--14, ashall--16, 15h, heringer--17, 12bwh, magee--19, vogl--19}). Comparing these parameters against explosion model predictions can hence provide constraints on the explosion physics for individual objects or samples, however this empirical process is also challenging.

\par

Determining which set of model parameters provide the best agreement is non-trivial. The most common approach is through visual inspection, in which input parameters are tweaked manually and the quality of the fit is determined purely through qualitative analysis. This makes it impossible to quantify the goodness of fit and experienced modellers may even disagree on which features are most important to reproduce, due to their own experience or biases. Furthermore, it is also not possible to determine whether the selected parameters represent a global or local minimum, as this would require computation of many thousands of spectra covering a huge parameter space that would each require visual inspection against the data. 

\par

Recent attempts have been made to provide more quantitative measures of the quality of fit and have shown promise \citep{ogawa--23}. At least some visual inspection and fine-tuning are still required however and fits can be limited by the model spectra available. Incorporating machine learning techniques into spectroscopic analysis could overcome some of the primary limitations, including the computational cost associated with fitting large numbers of parameters. \cite{chen--20} develop neural networks trained on \textsc{tardis} \citep{tardis} radiative transfer simulations and designed for predicting the chemical composition of SNe~Ia around maximum light. \cite{chen--24} apply this to estimate the \nickel abundance of SNe~Ia and investigate how this correlates with observed light curve properties. \cite{kerzendorf--21} present \textsc{dalek}, which is designed to emulate \textsc{tardis} simulations and predict model spectra with a speed up of $\gtrsim$10\,000. \cite{obrien--21} use \textsc{dalek} to quantitatively fit a spectrum of SN~2002bo, a nomal SN~Ia, approximately one week before maximum light and determine the chemical composition required to reproduce the observed spectral features. By comparing their inferred ejecta composition to explosion model predictions, they argue in favour of an explosion scenario containing a detonation for SN~2002bo. \cite{obrien--21} also find comparable results to previous studies without the need for manual and time-intensive tuning of model parameters, demonstrating the considerable power of this automated approach. \cite{obrien--23} extend this analysis to further apply \textsc{dalek} to model spectra of a sample of normal and 91T-like SNe~Ia \citep{91t--like} using customised ejecta profiles and investigate differences in their physical properties.

\par

Following from previous works, here we present \texttt{riddler}, a framework for automated, quantitative, and simultaneous fitting of full spectral time-series of SNe~Ia up to shortly after maximum light. Similar to previous studies \citep{kerzendorf--21, obrien--21, obrien--23} we begin by training a series of neural networks to emulate \textsc{tardis} radiative transfer simulations. As in \cite{obrien--21, obrien--23}, we then define a likelihood function and use nested sampling implemented in \textsc{ultranest} \citep{ultranest} to perform quantitative comparisons between our emulated model spectra and observed spectra of SNe~Ia. A key difference compared to previous works is that our training models cover a significantly larger time range and therefore rather than focus on indivdual spectra, our fitting approach has been extended to include multiple spectra at different epochs simultaneously. In addition, previous works have focused on inferring custom ejecta models for each SN. Here, we use realistic ejecta structures that are based on explosion simulations, allowing us to directly link the observed spectra back to predictions from explosion models and quantitatively determine the relative likelihood for the different ejecta structures predicted by the explosion models. Using \texttt{riddler}, we are able to generate large numbers of model spectra with significantly reduced computational cost, enabling a detailed investigation of different metrics by which the goodness of fit may be quantified. In Sect.~\ref{sect:training} we discuss our approach to generating a training dataset for our neural networks, which are described in Sect.~\ref{sect:nn}. In Sect.~\ref{sect:fitting} we discuss our approach to fitting spectra. Section~\ref{sect:observations} demonstrates \texttt{riddler} applied to spectra of SNe~Ia, while Sect.~\ref{sect:weightings} discusses the impact of different weighting schemes when calculating the likelihood. Finally, we discuss our results in Sect.~\ref{sect:discussion} and present our conclusions in Sect.~\ref{sect:conclusions}. \texttt{riddler} is publicly available on GitHub\footnote{\href{https://github.com/MarkMageeAstro/riddler}{https://github.com/MarkMageeAstro/riddler}}.

%

\section{Training models}
\label{sect:training}

\begin{figure*}
\centering
\includegraphics[width=\textwidth]{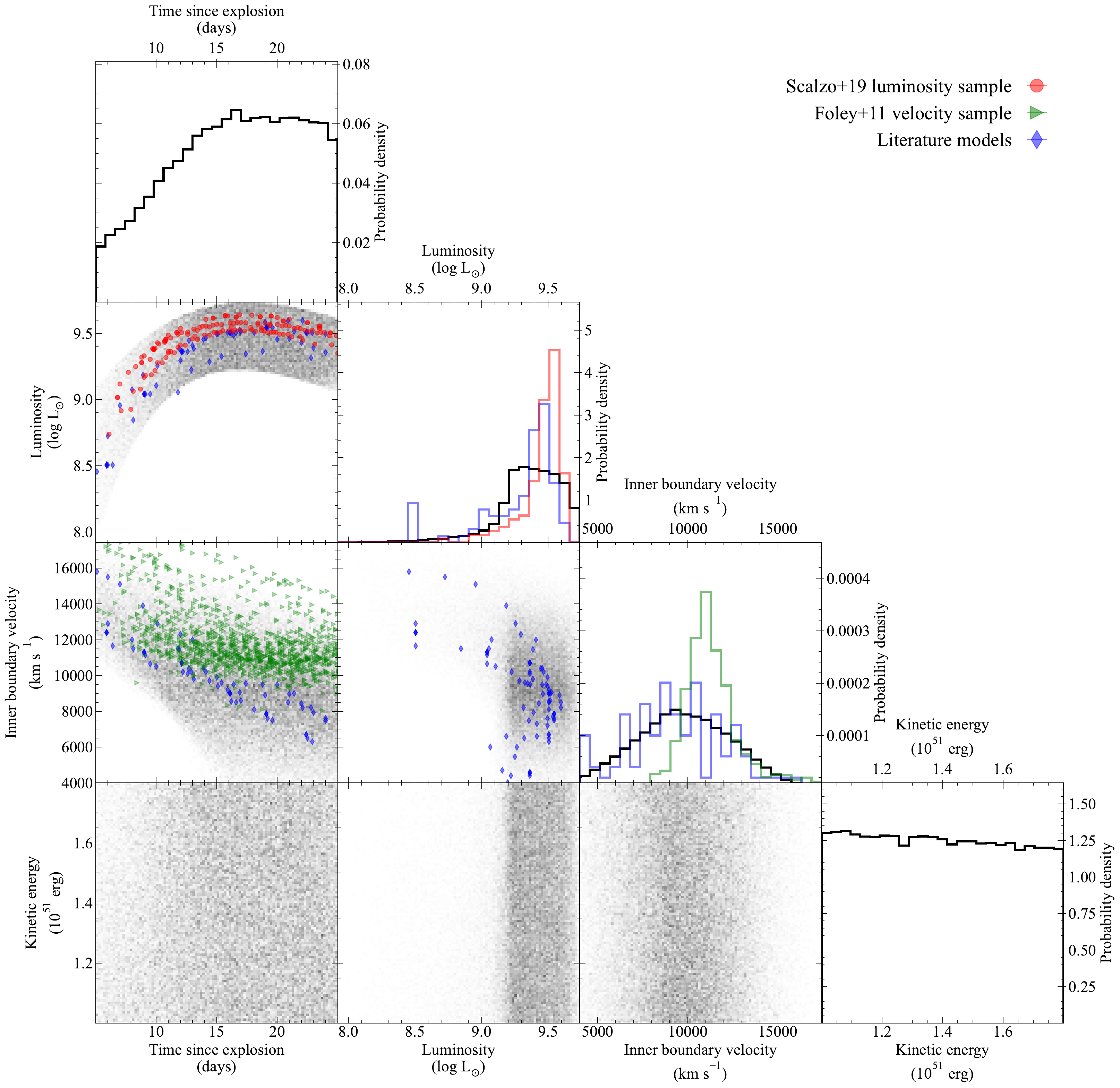}
\caption{Distributions of the sampled input parameters used by our neural networks are shown in black. Bolometric luminosities for a sample of SNe~Ia calculated by \citealt{scalzo--19} are shown in red. In green we show \sistff\, velocities calculated by \citealt{foley--11}. Literature models discussed in the text are shown in blue.}
\label{fig:params_dists}
\centering
\end{figure*}

The purpose of this work is to investigate the early, optical spectral evolution of SNe~Ia and investigate metrics through which to quantitatively determine the best-fitting ejecta structure predicted by different explosion models. For our spectral emulator neural networks, we therefore require a representative set of spectra that may be used for training. All model spectra used for training neural networks in this work were calculated with \textsc{tardis} \citep{tardis}. \textsc{tardis} is an open-source, one-dimensional Monte Carlo radiative transfer code designed to rapidly produce model spectra based on a given set of input parameters defined by the user. The flexibility and speed of \textsc{tardis}, in combination with the appropriate level of physics included, means that it is ideally suited for producing the tens of thousands of spectra necessary for training neural networks. In Sect.~\ref{sect:training_set} we discuss generating the \textsc{tardis} training set used for our neural networks, while in Sect.~\ref{sect:processing} we discuss the pre-processing applied before training the neural networks. The distributions of training parameters are shown in Fig.~\ref{fig:params_dists}.

\par
\subsection{Generating the training set}
\label{sect:training_set}

Each \textsc{tardis} simulation for producing a SN Ia spectrum requires a number of input parameters: time since explosion, luminosity, inner boundary velocity, and the density and composition of the ejecta. With our neural networks we wish to cover a large parameter space, such that our models may be used to fit observations of SNe~Ia at most epochs of interest. To determine a suitable range for each input parameter we use either samples of observed SNe~Ia or predictions from theoretical explosion models.

\par

We uniformly sample the time since explosion between 5~--~25\,d. We select a lower time boundary of 5\,d due to the increased computational cost in calculating models at very early times (as a result of the high density), while an upper boundary of 25\,d is selected due to the photospheric approximation made by \textsc{tardis}. \textsc{tardis} assumes a sharp inner boundary separating optically thick and thin regions, which limits its validity to times up to shortly after maximum light (see Sect.~\ref{sect:limitations}). For $\sim$10\% of models, we find that our \textsc{tardis} simulations did not converge after 20 iterations and therefore exclude them from our training datasets. These models were typically at early times and with high densities at the inner boundary. This results in a somewhat skewed distribution for the time since explosion, rather than uniform.

\par

Given that our sampled times since explosion cover a wide range of values, uniform sampling of the SN luminosity between fixed upper and lower boundaries would result in many models that are over/under-luminous at early/late-times compared to observations of SNe~Ia and hence are unphysical. Therefore, we adopt an empirical approach to determine the appropriate prior distribution for our input luminosity as a function of time since explosion. \cite{scalzo--19} constructed bolometric light curves for a sample of low-redshift, well-observed SNe~Ia observed by the Carnegie Supernova Project (CSP, \citealt{csp--survey}). 
Using these light curves we set an upper and lower limit on the bolometric luminosity as a function of time based on polynomial fits to the sample. Here we assume a typical rise time of 18.5\,d  \citep{ganeshalingam--2011, miller--20a}. We increase the upper limit on this luminosity range by a factor of 1.25, as \textsc{tardis} typically over-estimates the luminosity. Similarly, we decrease the lower limit by a factor of 0.5 to allow our models to fit fainter SNe than those included by \cite{scalzo--19}. For each training model, the luminosity is then uniformly sampled within the appropriate boundaries for the previously selected time since explosion. The distribution of luminosities used for our training set is shown in Fig.~\ref{fig:params_dists}, along with the observed SNe~Ia luminosities calculated by \cite{scalzo--19}. Figure~\ref{fig:params_dists} also shows that our expanded sample region covers the range of luminosities used by a selection of literature models collected from the following sources: \cite{stehle--05, mazzali--08, tanaka--11, ashall--14, mazzali--14, heringer--17, ashall--18, heringer--19}.

\par

\begin{figure}
\centering
\includegraphics[width=\columnwidth]{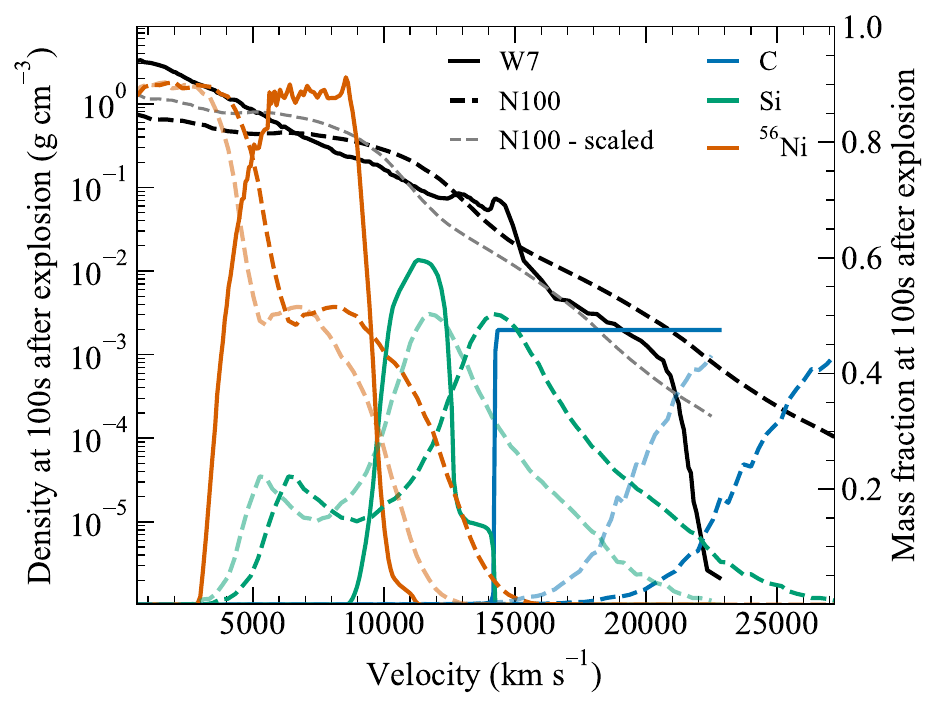}
\caption{Density and representative mass fractions for the W7 and N100 models at 100s after explosion. A scaled N100 density and composition profile assuming $KE' = 1.0\times10^{51}$~erg (Eqns.~\ref{eq:velocity_scale}~\&~\ref{eq:density_scale}) is shown as shaded lines.}
\label{fig:density_comp}
\centering
\end{figure}

For the inner boundary velocity we again adopt an empirical approach. We use the \ion{Si}{ii} velocities published by \cite{foley--11} and \cite{maguire--14} for samples of low-redshift SNe~Ia to estimate an appropriate sampling range as a function of time. By analysing the velocity evolution of their sample, \cite{foley--11} find that the \sistff\, velocity $v$ (in units of 10$^{3}$~km~s$^{-1}$) at a time relative to maximum light $t$ (in days) can be approximated with the following functional form:
\begin{equation}
\label{eq:velocity}
    v \left( v_0, t \right) = v_0 \left( 1 - 0.0322t \right) - 0.285t,
\end{equation}
where $v_0$ represents the \sistff\, velocity at maximum light. Using the previously selected time for each training model, and again assuming an 18.5\,d rise time, we uniformly sample between the appropriate upper and lower boundaries at that time, which are defined assuming $v_0$ velocities of 15\,000~km~s$^{-1}$ and 9\,000~km~s$^{-1}$ respectively. Again, we increase the upper limit of our sample range by a factor of 1.25. The model photospheric velocity in radiative transfer simulations is highly correlated with, but not necessarily equal to, the \ion{Si}{ii}~$\lambda$6\,355 velocity (Fig.~\ref{fig:params_dists}; c.f \citealt{parrent--2012, mazzali--14}). To account for differences between the observed \ion{Si}{ii}~$\lambda$6\,355 velocity and the model photospheric velocity we subtract an additional offset uniformly sampled between 0 -- 6\,000~km~s$^{-1}$. Figure~\ref{fig:params_dists} shows the velocities used for our training set along with photospheric velocities from a sample of literature models and \ion{Si}{ii}~$\lambda$6\,355 velocities for SNe~Ia measured by \cite{foley--11}. As shown by Fig.~\ref{fig:params_dists}, around maximum light the inner boundary velocity is typically lower than the observed \ion{Si}{ii}~$\lambda$6\,355 velocity.

\par

Unlike the training models used in previous works \citep{chen--20, obrien--21}, the density and composition of our training models does not follow an empirical approach and instead are more directly related to predictions from explosion models. For this initial, demonstrative study we limit our training datasets to compositions based on two well-studied explosion models: W7 \citep{nomoto-w7} and N100 \citep{seitenzahl--13, hesma}. The W7 model is a one-dimensional, parameterised pure deflagration explosion with a deflagration speed such that the resulting composition and spectra closely match observations of SNe~Ia around maximum light. The N100 model however is a deflagration-to-detonation transition explosion calculated via a fully realistic multi-dimensional simulation. Here we use the angle-averaged N100 model provided by HESMA\footnote{\href{https://hesma.h-its.org/}{https://hesma.h-its.org/}} \citep{hesma}. In Fig.~\ref{fig:density_comp} we show density profiles for both the W7 and N100 models, along with mass fractions of C, Si, and $^{56}$Ni. For each training set we assume the composition is fixed and taken directly from the respective explosion model. While both models do show good agreement with spectra of SNe~Ia around maximum light, neither provides perfect agreement \citep{branch--85, stehle--05, sim--13, mazzali--14}. These limitations will also be reflected in our training datasets and we therefore do not expect that either model will provide perfect agreement with observations. Nevertheless, by taking compositions directly from existing explosion models we avoid scenarios where the best-fitting model determined via fitting contains an unrealistic structure, which may arise from fitting the abundance of each element independently.

\par

Given that families of explosion models can generally produce similar structures with different kinetic energies (e.g. \citealt{seitenzahl--13}), we also allow for some variation in the ejecta structure of our training models by scaling to higher or lower kinetic energies \citep{hachinger--09, ashall--16}. The velocity and density of each cell in the model are scaled to a new kinetic energy $KE'$ and ejecta mass $M'$. The updated velocities $v'$ and densities $\rho'$ are given by:
\begin{equation}
\label{eq:velocity_scale}
    v' = v_{exp} \left( \frac{KE'}{KE_{exp}} \right)^{\frac{1}{2}} \left( \frac{M'}{M_{exp}} \right)^{-\frac{1}{2}},
\end{equation}
\begin{equation}
\label{eq:density_scale}
    \rho' = \rho_{exp} \left( \frac{KE'}{KE_{exp}} \right)^{-\frac{3}{2}} \left( \frac{M'}{M_{exp}} \right)^{\frac{5}{2}},
\end{equation}
where $KE_{exp}$ and $M_{exp}$ are the kinetic energies and ejecta masses of the input explosion models, in this case W7 and N100. For the current work, as we are focused only on Chandrasekhar mass explosions, we assume the ejecta mass is fixed and only allow for variation in the kinetic energy. We uniformly sample the kinetic energy between $10^{51}$~--~$10^{51.26}$~erg, which covers the range of predictions from Chandrasekhar mass explosion models\footnote{We note that \textsc{tardis} simulations do not include any information about the ejecta below the inner boundary velocity and hence cannot be used to constrain the entire ejecta. Therefore we do not claim to measure the true kinetic energy, or indeed the full structure, of the ejecta via our fitting method. The parameterisation used here is simply a useful means for controlling the ejecta structure in our models and introducing some variation beyond the two explosion models considered.} \citep{nomoto-w7, sim--13}.

\par

In summary, we generate two complete training datasets based on the W7 and N100 explosion models. For both datasets we use the same input parameters, which are defined as time since explosion, luminosity, inner boundary velocity, and kinetic energy, and the same randomly selected values for each input parameter. Compared to the training sets presented by \cite{chen--20} and \cite{obrien--21}, our models cover a wider range of times and luminosities, making it possible to fit a sequence of spectra for a single SN~Ia and determine the changes in the best fitting parameters. In addition to the ranges previously specified, we apply further criteria to ensure that our randomly selected parameter values represent physical and useful models. We set a minimum inner boundary velocity of 4\,000~km~s$^{-1}$ and reject models where the inner boundary is \textless2\,500~km~s$^{-1}$ below the maximum velocity of the ejecta. We randomly generate 120\,000 models that meet these requirements for both our W7 and N100 training datasets. Of these, we use 100\,000 to train the neural networks and 20\,000 test models to determine the typical accuracy of the neural networks (see Sect.~\ref{sect:appdx:nn_accuracy}). The choice of appropriate prior space for our input parameters does require some manual selection and will significantly impact the determination of the overall best-fitting explosion model. We stress that we are only able to comment on which model provides a better match to the data within the boundaries set by our priors. In Sect.~\ref{sect:limitations} we discuss the limitations of our training datasets and future improvements.

\subsection{Pre-processing}
\label{sect:processing}

\begin{figure*}
\centering
\includegraphics[width=\textwidth]{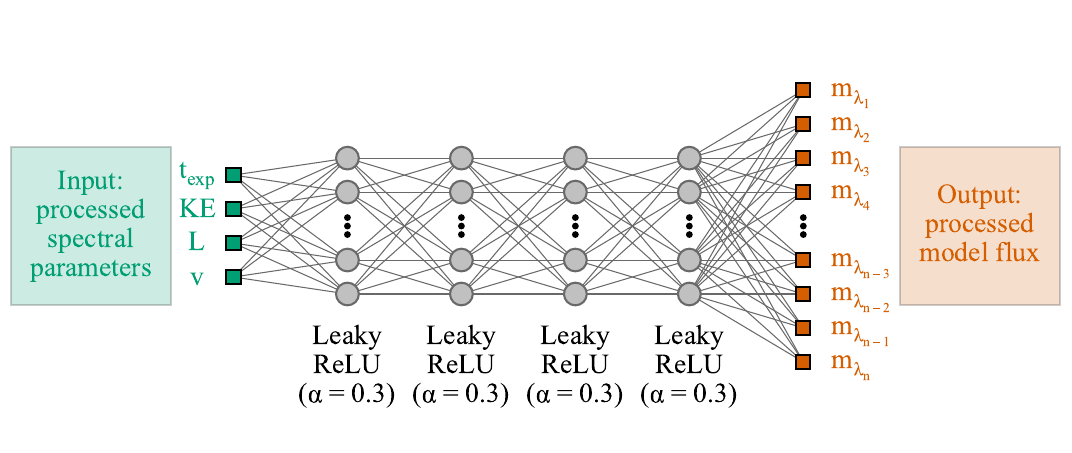}
\caption{Schematic diagram showing an example neural network architecture with four hidden layers. As input the neural network takes the processed spectral parameters defining our \textsc{tardis} models, which are the time since explosion ($t_{exp}$), the kinetic energy of the ejecta (KE), the luminosity of the spectrum ($L$), and the inner boundary velocity ($v$). Each neuron uses a leaky-ReLU activation function in which the gradient for negative values is set to $\alpha = 0.3$. The outputs of the neural network are the processed flux values within each wavelength bin, $m_{\lambda_{n}}$. }
\label{fig:nn_architecture}
\centering
\end{figure*}

The models developed for training as part of this work cover a wide range of luminosities within each wavelength bin and input parameters. Therefore, they require pre-processing before they can be used effectively by the neural network. Our first step is to rebin all of the model spectra. We chose 1\,000 log-spaced wavelength bins in the range 3\,000 -- 9\,000~\AA. To allow our neural network to be used for fitting a wide range of SN~Ia spectra, with different signal-to-noise ratios, we follow \cite{chen--20} and apply smoothing to each spectrum using a Savitzky-Golay filter (order 2, window 25). 

\par

We also experimented with data augmentation to boost the size of our training dataset, as was done by \cite{chen--20}. For this augmented training data, each spectrum was included ten times with different levels of smoothing applied, resulting in ten slightly different spectra for a given set of input parameters. This was designed to approximately mimic different signal-to-noise ratios. Despite the increase in the size of the available training dataset, we found the overall performance of the neural networks decreased following data augmentation with this method. This likely arose from the different levels of smoothing resulting in varying degrees of blending for spectral features and noise within the training data inhibiting the ability of the neural networks to learn effectively. We therefore do not include additional data augmentation during training, but future work will explore alternative augmentation methods. 

\par

After rebinning and smoothing, the luminosity within each wavelength bin is processed by first taking log$_{10}$ and transforming using the \textsc{standard} scaler implemented in \textsc{sklearn}. Each of the input parameters used for generating our \textsc{tardis} spectra are processed in the same manner.

%

\section{Neural networks}
\label{sect:nn}

All neural networks were trained using \textsc{tensorflow} \citep{tensorflow} and \textsc{keras} \citep{keras}. Extensive hyperparameter testing and tuning was performed both manually and with \textsc{optuna} \citep{optuna}. For both the W7 and N100 models, we train a final set of 20 neural networks each that are discussed throughout this work. Figure~\ref{fig:nn_architecture} shows an example of the neural network architecture used. Table~\ref{tab:nn_hyper} in the Appendix gives the hyperparameters of each neural network, along with the median mean fractional error and median maximum fractional error across the 20\,000 W7 and N100 test models. 

\par

The input layer of each neural network is defined by four neurons, corresponding to the four (processed) parameters used to generate each \textsc{tardis} spectrum (time since explosion, luminosity, inner boundary velocity, and kinetic energy). All neural networks are trained on the same set of 100\,000 input parameters with the same validation split of 20\%. The number of subsequent layers and neurons per layer are given in Table~\ref{tab:nn_hyper}. We do not include batch normalisation after any layer, as this was found to decrease the overall performance of the neural networks. The output layer is defined by 1\,000 neurons and corresponds to the (processed) flux in each of the 1\,000 wavelength bins. In general, we found that the choice of optimiser had little impact on the performance and therefore chose the Nadam optimiser for all neural networks. In addition, we found the leaky-ReLU \citep{lrelu} activation function and mean squared error loss function provided the best performance and therefore were used for all neural networks during training. 

\par

Each neural network was trained for a total of 50\,000 epochs on three Nvidia Quadro RTX 6\,000 GPUs. To prevent over-fitting, we monitor the performance of the neural networks on the validation data and save the epoch with the best validation performance. In Sect.~\ref{sect:appdx:nn_accuracy} in the Appendix, we discuss the accuracy of our neural networks in more detail, but we find typical accuracies of up to a few percent (Table~\ref{tab:nn_hyper}).

%

\section{Spectral sequence fitting method}
\label{sect:fitting}

In the following section, we discuss our approach to fitting multiple spectra of SNe~Ia simultaneously. Using the neural networks described in Sect.~\ref{sect:nn}, we aim to determine the best-fitting model input parameters for each observed spectrum. In other words, the set of input parameters $\theta$ with the highest likelihood given the observations $o$, $\mathcal{L}(\theta|o)$. We also aim to quantitatively determine the relative likelihood of the W7 and N100 models for a given SN~Ia by comparing the evidence $\mathcal{Z}$ for each model \citep{thrane--20} assuming our goodness-of-fit metric and priors. In Bayesian inference, the evidence is given by the likelihood function marginalised over the prior distribution $\pi(\theta)$:
\begin{equation}
\label{eqn:evidence}
    \mathcal{Z} = \int \mathcal{L}(\theta|o) \pi(\theta) d\theta,
\end{equation}
which is calculated for both the W7 and N100 models based on the total likelihood across all spectra included in the fit.

\par

The prior distributions, $\pi(\theta)$, for our input parameters used during fitting are the same as those in Sect.~\ref{sect:training_set}. As discussed in Sect.~\ref{sect:training_set}, the input parameters used to generate our \textsc{tardis} spectra are given by the time since explosion, luminosity ($L$), inner boundary velocity ($v$), and the density and composition of the ejecta. For the current work, the density and composition are taken from either the W7 or N100 model, with some variation given by the kinetic energy (KE). We note that as we do allow for variation in the kinetic energy, the emulated models are not strictly the exact predictions from either W7 or N100. We therefore refer to the resulting best-fit models as either W7- or N100-like. The explosion epoch, $t_{exp}$, and KE are both fixed for all spectra in the sequence. This results in a set of $2 + 2N$ parameters defining the spectral sequence, $\theta$~=~\{$t_{exp}$, KE, $L_1$, $v_1$, ..., $L_N$, $v_N$\}, where $N$ is the number of spectra included in the fit. To reduce degeneracy between the free parameters, we add an additional constraint to ensure that the inner boundary velocity decreases for later spectra, $v_i \geq v_{i+1}$, which is typically observed in SNe~Ia \citep{foley--11, maguire--14} and used in spectral modelling \citep{stehle--05}. In Sect.~\ref{sect:limitations}, we discuss the impact of this assumption.

\par

Fits are performed with \textsc{ultranest} \citep{ultranest}. The benefit of nested sampling routines such as \textsc{ultranest} is that they can be used to simultaneously infer posterior distributions and calculate the model evidence. For each set of input parameters $\theta$, we use a subset of the neural networks described in Sect.~\ref{sect:nn} to generate synthetic spectra covering the spectral sequence. We assume a Gaussian likelihood function for each input spectrum $i$ given by,
\begin{equation}
\label{eqn:likelihood}
    \ln \mathcal{L}_i(\theta|o) \propto -\frac{1}{2}\sum_{\lambda} w_{\lambda} \left[ \left( \frac{m_{\lambda}(\theta) - o_{\lambda}}{s_{\lambda}(\theta)} \right)^2 + \ln (2 \pi s_{\lambda}(\theta)^2 ) \right],
\end{equation}
where 
\begin{equation}
\label{eqn:uncertainty}
    s_{\lambda}(\theta)^2 = \sigma_{o,\lambda}^2 + f^2 m_{\lambda}(\theta)^2 + \Delta_{\lambda}(t, v)^2 m_{\lambda}(\theta)^2.
\end{equation}
Here, $o_{\lambda}$ and $\sigma_{o,\lambda}$ give the observed flux and uncertainty, $m_{\lambda}(\theta)$ gives the model flux for a given set of input parameters $\theta$, $\Delta_{\lambda}(t, v)$ gives the fractional uncertainty of the prediction for the input time since explosion $t$ and inner boundary velocity $v$, $f$ is a nuisance parameter included to account for underestimated uncertainties, and $w$ is a weighting parameter. The integration is performed over all wavelengths $\lambda$. In cases where the observed flux uncertainty is unavailable we assume an uncertainty of 2\%. Sect.~\ref{sect:appdx:nn_accuracy} discusses the uncertainty of our neural network predictions ($\Delta_{\lambda}(t, v)$) in more detail, however we note that this is typically of order a few percent. The parameter $f$ is included as an additional source of uncertainty to account for systematic differences between off-the-shelf explosion model predictions and individual SNe~Ia. We note that these systematic differences between explosion model spectra and observed SNe~Ia are typically $\gtrsim$20\% \citep{kerzendorf--21}. Therefore, without this additional parameter, the posteriors give unreasonably well-constrained parameters. This arises from the fact that none of the models constitute a `good' fit based on the $\chi^2$ value and therefore, small deviations away from the best-fitting model result in large changes in the $\chi^2$ value and only a narrow range of acceptable parameters. The parameter $w_{\lambda}$ allows for different relative weighting of individual features or wavelengths. By default, we make the simplest assumption of no weighting applied during fitting. For the initial work presented here, we also explore the impact of excluding different wavelengths during fitting in Sect.~\ref{sect:weightings}. We note that \texttt{riddler} allows for the inclusion of any arbitrary weighting scheme implemented by the user.

\par

Fits are performed independently with each of the top six neural networks as determined by their mean and maximum fractional errors (see Table~\ref{tab:nn_hyper} in the Appendix). We combine their posteriors determined by \textsc{ultranest} to estimate the overall best-fitting parameters. This approach helps to overcome limitations associated with the accuracy of any given neural network and provides a more reliable estimate for the total uncertainty of the model parameters, including systematic uncertainties associated with the different neural network architectures. For the W7 model, the best performing neural networks used throughout the rest of this work are NNs 2, 3, 4, 16, 17, \& 18. For the N100 model, these are NNs 8, 13, 14, 16, 17, \& 18. To compare the relative likelihoods of two models (in this case either W7- or N100-like ejecta structures) we use the evidences calculated by \textsc{ultranest} and the Bayes factor given by 
\begin{equation}
\label{eqn:bf}
    \ln \mathrm{BF} = \ln \mathcal{Z}_1 - \ln \mathcal{Z}_2,
\end{equation}
where $\mathcal{Z}_{1, 2}$ refer to the evidence for either W7- or N100-like models. In this case, a Bayes factor \textgreater1 indicates a preference for model 1 relative to model 2.

%

\section{Application}
\label{sect:observations}

Having described the neural networks designed to emulate \textsc{tardis} simulations and our assumed likelihood function, we now apply \texttt{riddler} to fitting spectra. We begin by applying \texttt{riddler} to a model spectrum, for which the true input parameters are known, and then to observed SNe~Ia.

\begin{figure*}
\centering
\includegraphics[width=\textwidth]{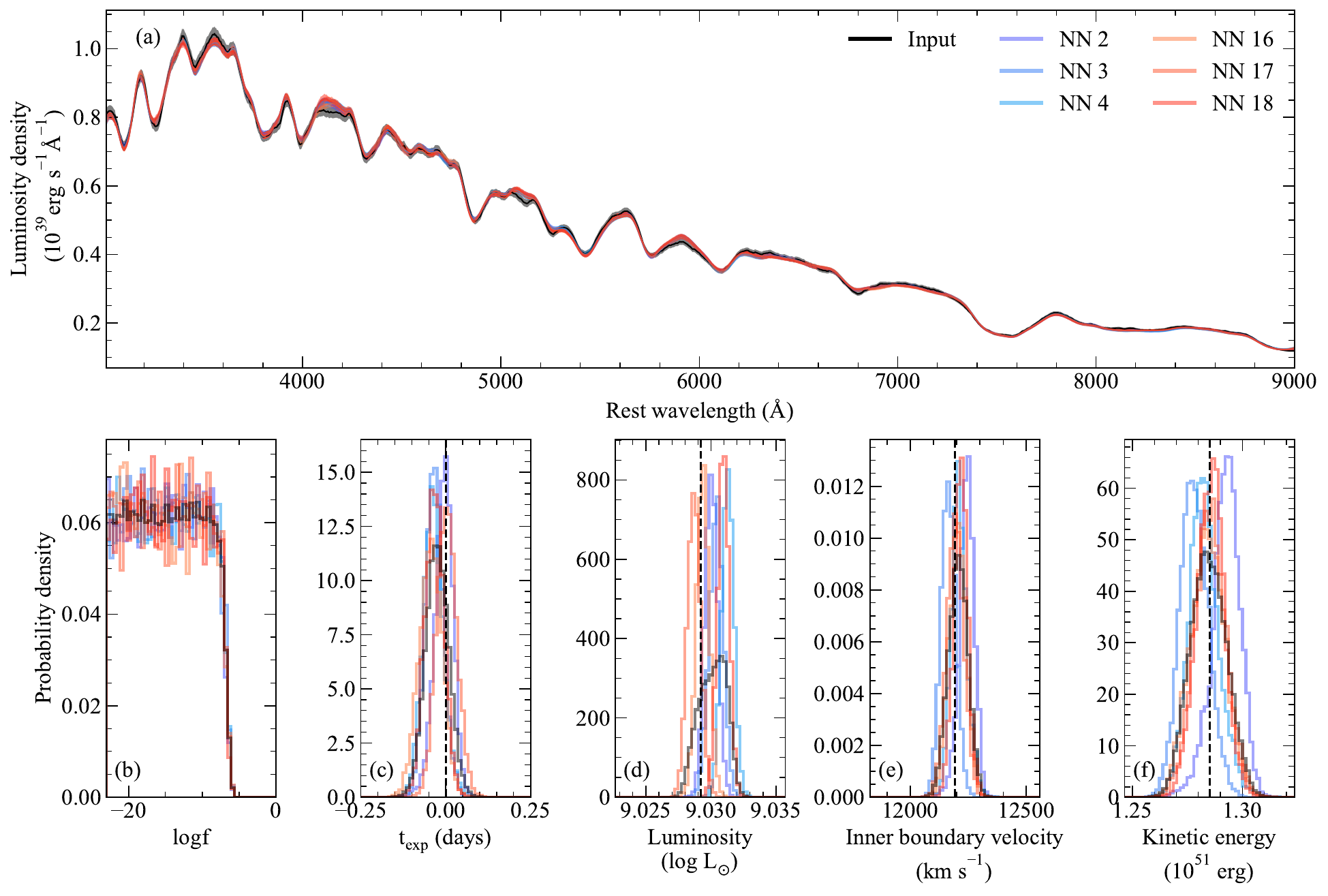}
\caption{\textit{Panel a:} Comparison between our input spectrum (black) and the neural network reconstructions of the best-fitting parameters based on our \textsc{ultranest} fits. The grey shaded region shows the assumed 2\% uncertainty on the model flux, while coloured shaded regions show the model flux error, $\Delta_{\lambda}$. \textit{Panels b -- f:} Posterior distributions of fitting parameters determined by \textsc{ultranest}. The posterior distribution for each neural network is shown as a coloured line, while the total posterior across all neural networks is shown as a grey line. In \textit{Panels c -- f} we also denote the true input value as a vertical dahsed line. }
\label{fig:w7_fit_test}
\centering
\end{figure*}

\begin{table*}
\centering
\caption{Best-fitting parameters, likelihoods, and evidences for SN~2011fe and SN~2013dy with either W7 or N100 models assuming uniform wavelength weighting}\tabularnewline
\label{tab:spec_log}\tabularnewline
\begin{tabular}{lllllllll}\hline
\hline
\tabularnewline[-0.25cm]
& & & \multicolumn{3}{c}{W7} & \multicolumn{3}{c}{N100} \tabularnewline
Date  & MJD & Phase  &  Luminosity        & Inner boundary  & $\ln \mathcal{L}(\theta|o)$ &  Luminosity        & Inner boundary  & $\ln \mathcal{L}(\theta|o)$ \tabularnewline
      &     & (days) &  (log $L_{\odot}$) & (km~s$^{-1}$)   &                         &  (log $L_{\odot}$) & (km~s$^{-1}$)   &                         \tabularnewline
\hline
\multicolumn{9}{c}{SN~2011fe} \tabularnewline
\hline
2011 Aug. 28 & 5\,5801.17 & \phn$-$13.1   &  ${8.56^{8.58}_{8.54} }$ & ${12221^{12598}_{11623} }$ 		& ${-88349_{-88357}^{-88330} }$ & ${8.56^{8.58}_{8.55} }$ & ${13987^{14348}_{13583} }$  	& ${-88263_{-88273}^{-88252} }$ \vspace{0.2cm}	    \tabularnewline 
2011 Aug. 31 & 5\,5804.25 & \phn$-$10.1   &  ${9.11^{9.12}_{9.09} }$ & ${12181^{12589}_{11591} }$ 		& ${-89153_{-89158}^{-89148} }$ & ${9.10^{9.11}_{9.08} }$ & ${12359^{13013}_{11828} }$  	& ${-89197_{-89200}^{-89194} }$ \vspace{0.2cm}	   \tabularnewline 
2011 Sep. 03 & 5\,5807.38 & \phn$-$6.9    &  ${9.40^{9.42}_{9.38} }$ & \phn{}${6740^{6956}_{6671} }$ 	& ${-89730_{-89735}^{-89727} }$ & ${9.39^{9.41}_{9.38} }$ & ${11267^{11995}_{10829} }$ 		& ${-89785_{-89788}^{-89783} }$	\vspace{0.2cm}	    \tabularnewline 
2011 Sep. 07 & 5\,5811.37 & \phn$-$2.9    &  ${9.53^{9.54}_{9.51} }$ & \phn{}${5730^{6185}_{5330} }$    & ${-90022_{-90024}^{-90020} }$ & ${9.52^{9.54}_{9.51} }$ & \phn{}${7225^{7738}_{6837} }$ 	& ${-90248_{-90253}^{-90239} }$ \vspace{0.2cm}	    \tabularnewline 
2011 Sep. 10 & 5\,5814.39 & \phn$+$0.1    &  ${9.54^{9.56}_{9.52} }$ & \phn{}${4840^{5320}_{4350} }$	& ${-90225_{-90227}^{-90222} }$ & ${9.52^{9.54}_{9.50} }$ & \phn{}${4035^{4340}_{4000} }$	& ${-90288_{-90295}^{-90280} }$ \vspace{0.2cm}	   \tabularnewline 
2011 Sep. 13 & 5\,5817.67 & \phn$+$3.4    &  ${9.50^{9.53}_{9.48} }$ & \phn{}${4033^{4338}_{4000} }$	& ${-90520_{-90525}^{-90514} }$ & ${9.49^{9.51}_{9.47} }$ & \phn{}${4011^{4249}_{4000} }$ 	& ${-90526_{-90538}^{-90517} }$ \vspace{0.2cm}	    \tabularnewline 
\hline
& & & & $\ln \mathcal{Z}$ & ${-538059_{-538082}^{-538032} }$ & & $\ln \mathcal{Z}$ & ${-538366_{-538377}^{-538351} }$  \tabularnewline
\hline\hline
\multicolumn{9}{c}{SN~2013dy} \tabularnewline
\hline
2013 Jul. 21 & 5\,6494.48 & \phn$-$6.6     & ${9.42^{9.44}_{9.40} }$ & \phn{}${8956^{9490}_{8404} }$ & ${-89846_{-89853}^{-89835} }$ & ${9.41^{9.45}_{9.39} }$ & ${11167^{13222}_{10325} }$ 		& ${-89850_{-89868}^{-89833} }$ \vspace{0.2cm}	    \tabularnewline 
2013 Jul. 25 & 5\,6498.60 & \phn$-$2.5     & ${9.50^{9.53}_{9.48} }$ & \phn{}${5307^{5519}_{5009} }$ & ${-89943_{-89948}^{-89936} }$ & ${9.48^{9.51}_{9.46} }$ & ${10422^{12155}_{9206} }$  		& ${-90287_{-90330}^{-90254} }$ \vspace{0.2cm}	    \tabularnewline 
2013 Jul. 27 & 5\,6500.32 & \phn$-$0.8     & ${9.51^{9.54}_{9.48} }$ & \phn{}${4465^{4613}_{4257} }$ & ${-89945_{-89954}^{-89939} }$ & ${9.48^{9.51}_{9.47} }$ & \phn{}${9961^{11743}_{8510} }$   & ${-90451_{-90474}^{-90432} }$ \vspace{0.2cm}	    \tabularnewline 
2013 Jul. 29 & 5\,6502.31 & \phn$+$1.2     & ${9.48^{9.50}_{9.46} }$ & \phn{}${4003^{4066}_{4000} }$ & ${-90012_{-90019}^{-90001} }$ & ${9.47^{9.50}_{9.45} }$ & \phn{}${9293^{11346}_{7572} }$   & ${-90559_{-90569}^{-90530} }$ \vspace{0.2cm}	    \tabularnewline 
2013 Aug. 01 & 5\,6505.57 & \phn$+$4.5     & ${9.41^{9.43}_{9.39} }$ & \phn{}${4001^{4026}_{4000} }$ & ${-90241_{-90250}^{-90232} }$ & ${9.42^{9.44}_{9.40} }$ & \phn{}${5363^{6417}_{4000} }$    & ${-90564_{-90596}^{-90462} }$ \vspace{0.2cm}	    \tabularnewline 
\hline
& & & & $\ln \mathcal{Z}$ & ${-450039_{-450050}^{-450024} }$ & & $\ln \mathcal{Z}$ & ${-451751_{-451762}^{-451739} }$  \tabularnewline
\hline
\hline
\multicolumn{9}{p{17cm}}{\textit{Note.} The best-fitting model parameters are determined based on the median of the total posterior across all of the neural networks included in the fits. Upper and lower limits are also based on the total posterior across all neural networks. We stress that these parameters assume uniform wavelength weighting, which has important limitations (see Sect.~\ref{sect:weightings}). Likelihoods and evidences are given by the mean with upper and lower limits determined from the minimum and maximum values, across all neural networks,.}  \tabularnewline
\end{tabular}
\end{table*}

\subsection{Test models}

Here we apply \texttt{riddler} to fitting model spectra generated using the same approach discussed in Sect.~\ref{sect:training_set} (but not seen during training) and demonstrate that it is able to accurately recover the parameters of the input model. During fitting, we assume a 2\% flux uncertainty to mimic real data and account for noise in the radiative transfer simulation. In Fig.~\ref{fig:w7_fit_test} we show an example fit applied to one of our W7 model spectra. We find comparable results when fitting our N100 spectra. 

\par

Figure~\ref{fig:w7_fit_test}(a) shows our input model spectrum compared against neural network reconstructions of the best-fitting parameters. The mean fractional errors of these spectra are $\sim$1.3\%. As shown by Fig.~\ref{fig:w7_fit_test} our best-fitting spectra reproduce the strongest spectral features in the input spectrum with the correct velocity, strength, and width. For weaker spectral features however, such as those at $\lambda \sim$4\,500\,\AA\, and $\sim$5\,150\,\AA, our best-fitting spectra struggle to reproduce the input model. We note that the inability to reproduce weak features is a systematic issue with our neural networks (Sect.\ref{sect:appdx:nn_accuracy}) and could result in absent or blended features in the best-fitting spectra. Our assumed 2\% flux uncertainty likely exacerbates the inability to fit weak features further as they can become lost in the spectrum uncertainty, but is nevertheless a better representation of fitting real data. Overall, our best-fitting models also reproduce the flux level and colour of the input spectrum, but there are some systematic differences. Around $\sim$3\,400 -- 3\,600\,\AA\, the neural networks predict systematically lower flux than in the input model, while around $\sim$4\,000 -- 4\,100\,\AA\, they predict systematically higher flux. In both cases however the neural network predictions are within our assumed 2\% uncertainty for the input model spectrum. 

\par

Figure~\ref{fig:w7_fit_test}(b) shows that the posterior distribution of $\log f$ is consistent with essentially no additional systematic uncertainty ($\lesssim0.5\%$), which is unsurprising given that the model was constructed in the same manner as the training dataset and with the same underlying ejecta structure. In Figs.~\ref{fig:w7_fit_test}(c) -- (f) we present posteriors for the best-fitting parameters of the input spectrum compared to the true value. While individual neural networks do show some variation in their posterior distributions, they all generally produce results that are consistent with each other and with the true input value. We note that in some cases the posterior means may be systematically offset from the input value. We therefore take the upper and lower limits of the full posterior across all neural networks as a conservative estimate of the total uncertainty. These limits are typically $\lesssim\pm5\%$ of the input value and include the input value itself.

\subsection{Observations}

Having shown that \texttt{riddler} is able to accurately recover the input parameters of our \textsc{tardis} model spectra, we now apply it to observations and estimate the best-fitting parameters for observed SNe~Ia. For this purpose, we use the well-observed SN~2011fe \citep{11fe--nature}, which was the subject of previous, detailed spectroscopic modelling by \cite{mazzali--14} and \cite{heringer--17}, and SN~2013dy \citep{zheng--13}. All spectra were obtained from WISeREP \citep{wiserep}. Both SNe are fit independently using our W7 and N100 models. No wavelength-dependent weighting (see Sect.~\ref{sect:weightings}) or additional flux scaling has been applied to these fits. Details of the spectra included during fitting are given in Table~\ref{tab:spec_log}, along with best-fitting values, and upper and lower limits for each input parameter. In addition, Table~\ref{tab:spec_log} also gives the mean, minimum, and maximum evidences (Eqn.~\ref{eqn:evidence}) for a given model based on the total likelihood across all spectra included in the fit and likelihoods (Eqn.~\ref{eqn:likelihood}) for a given model and individual spectra, as determined from our \textsc{ultranest} fits. 

\par

We again note that as we are using predictions from explosion models, and do not allow the structure or composition of our model ejecta to vary freely, we do not expect to find perfect agreement with the observations. Our approach with \texttt{riddler} however is able to find the best-fitting set of input parameters, including explosion epoch, for a given explosion model template assuming our likelihood function and prior distributions. We focus on quantifying the relative agreement between different models, based on the likelihoods and evidences.

\begin{figure*}
\centering
\includegraphics[width=\textwidth]{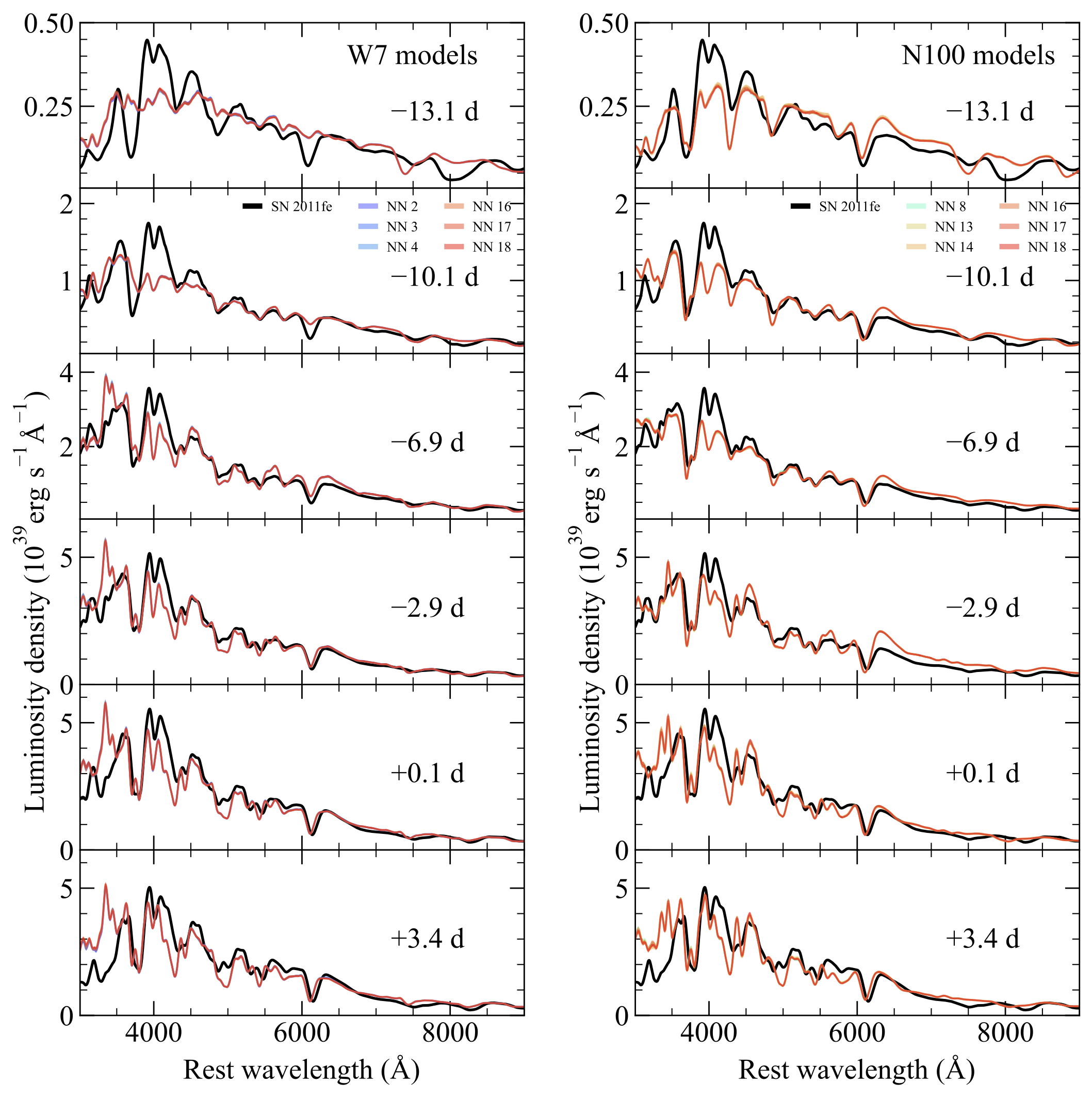}
\caption{Comparison between SN~2011fe and neural network reconstructions of the best-fitting W7 (\textit{left}) and N100 (\textit{right}) model spectra for each neural network assuming uniform wavelength weighting. Spectra are shown on an absolute luminosity scale with no additional scaling or offsets applied. Phases are given relative to $B$-band maximum.}
\label{fig:11fe_comp}
\centering
\end{figure*}

\begin{figure*}
\centering
\includegraphics[width=\textwidth]{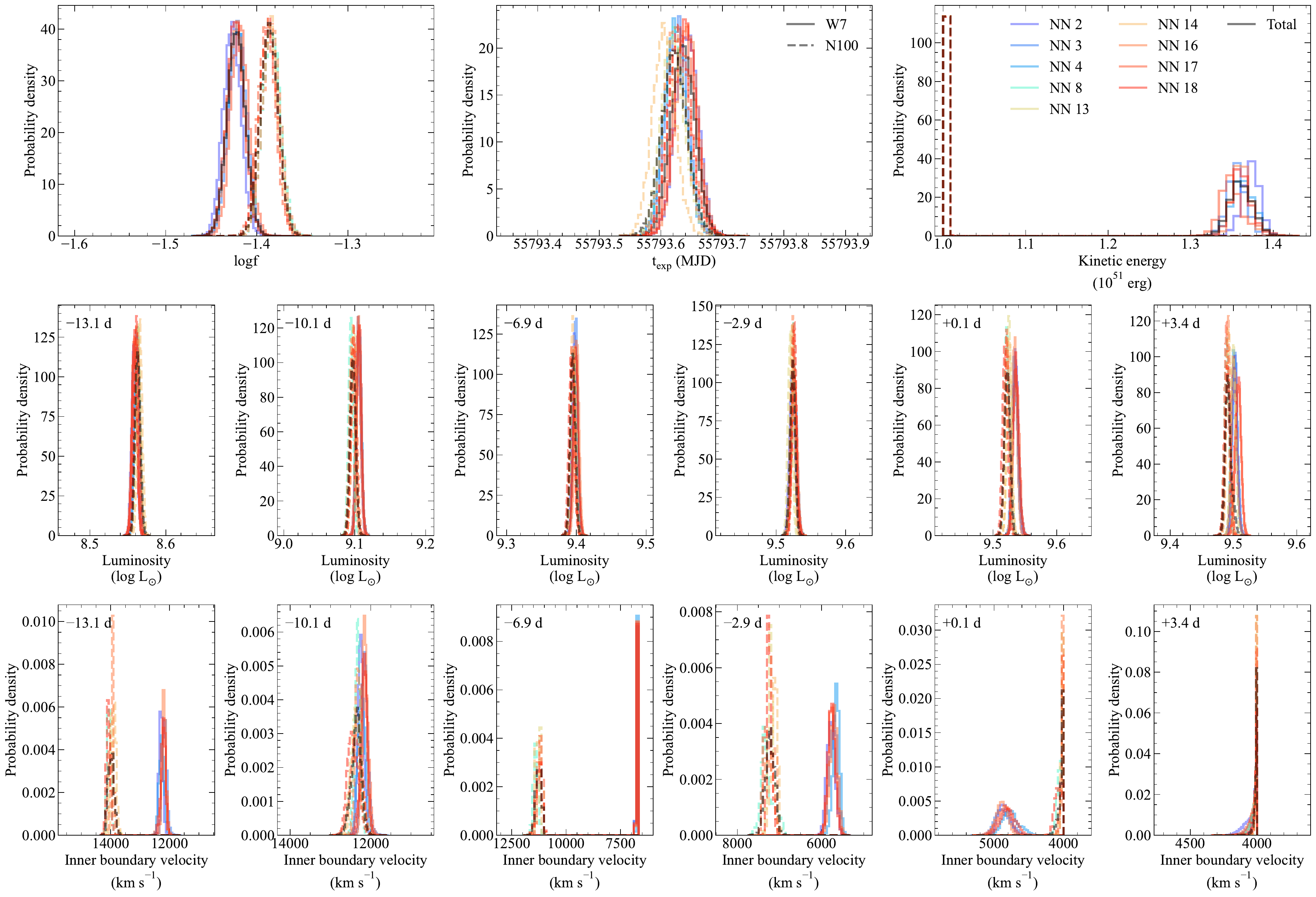}
\caption{Posterior distributions for best-fitting parameters to SN~2011fe using W7 and N100 models for each neural network assuming uniform wavelength weighting. Phases are given relative to $B$-band maximum.}
\label{fig:11fe_params}
\centering
\end{figure*}

\subsubsection{SN~2011fe}
\label{11fe}
For SN~2011fe, we use the HST spectra presented by \cite{mazzali--14} and assume a distance modulus of $\mu = 29.04$ mag and total extinction of $E(B-V) = 0.02$ mag \citep{mazzali--14}. Figure~\ref{fig:11fe_comp} shows the spectra reconstructed by our neural networks for the W7 and N100 models and their respective best-fitting parameters. The posterior distributions from each fit are shown in Fig.~\ref{fig:11fe_params}.

\par

From Fig.~\ref{fig:11fe_comp} it is clear that the earliest SN~2011fe spectrum is not reproduced by a W7-like ejecta as the models predict features that are significantly weaker than those observed. At this phase models with an N100-like ejecta produce features that more closely resemble SN~2011fe, although the overall spectral shapes show some differences and are redder than SN~2011fe. Fits with our N100 models favour kinetic energies towards the lower boundary of our input parameters ($\sim1.0 \times 10^{51}$~erg), which represents a $\sim$30\% decrease in the kinetic energy and results in scaled density profiles similar to that of W7 (Fig.~\ref{fig:density_comp}). Given that the best-fitting models approach the lower boundary, this would indicate that our input parameter space is not sampling the parameter range necessary to fully reproduce SN~2011fe, however further decreasing the kinetic energy may be unphysical. At this phase the ejecta above the photosphere in the W7-like models is dominated by unburned carbon and oxygen, while the N100-like models have more extended distributions of iron-group and intermediate-mass elements and hence show stronger spectral features. At $-10.1$\,d we again find that the N100-like models produce better agreement with SN~2011fe for some spectral features compared to the W7-like models. In particular, the N100-like models show stronger \sistff\, and \cahk that are more consistent with SN~2011fe, but over-predict the strength of the \ion{Si}{ii} and \ion{Fe}{iii} blend around $\sim$4\,800~\AA\, and near-ultraviolet (NUV) flux. Despite showing visually worse agreement with SN~2011fe spectral features, the W7-like models are favoured at this phase based on the likelihood, which is primarily driven by improved agreement with the continuum at longer wavelengths and the need for a higher systematic uncertainty ($f$) for the N100-like models. In addition, while the pseudo-equivalent widths of features for the N100-like models show better agreement with SN~2011fe, the lower overall flux around these wavelengths (relative to the observations) can lead to lower likelihoods. Conversely, the weaker features in the W7-like models can lead to higher likelihoods. This would indicate that additional weighting, potentially based on the overall strengths of the features, could be considered to avoid situations where spectra that do not produce features are favoured quantitatively \citep{ogawa--23}. By $-6.9$\,d W7-like models begin to show improved agreement with SN~2011fe and are generally able to match the \sistff, \sifnst, and \ion{S}{ii}-W features, and the complex silicon/iron blend around $\sim$5\,000~\AA. Again however we find that the N100-like models show better fits to these features, but the W7-like models have higher likelihoods due to their better agreement with the continuum at $\lambda \gtrsim 6\,500$~\AA.

\par

Around maximum light, the differences between our best-fitting W7-like and N100-like models become less pronounced. Both ejecta structures generally reproduce many of the spectral features, but systematically over-estimate the NUV flux and do not reproduce NUV spectral features. Our W7-like models also produce reasonable agreement with the \ion{Ca}{ii} NIR triplet, while our N100-like models are not able to match this feature. As with earlier epochs, the W7-like models are better able to match the flux at longer wavelengths and therefore produce higher likelihoods. We note however that for the spectrum at $+3.4$\,d, both models show overlapping likelihood ranges across all of the neural networks used here, indicating no strong preference for either model at this epoch.

\par

Figure~\ref{fig:11fe_params} shows that the W7-like and N100-like models have similar levels of systematic differences (parameterised by $\log f$, $\sim$23\%) between the models and observed spectra, although N100-like models are typically higher. Despite differences in the overall level of agreement, both sets of models predict a consistent explosion epoch of MJD = 5\,5793.6$\pm$0.1 and consistent luminosities for each spectrum. As previously mentioned, our N100-like model fits favour a density profile scaled to a lower kinetic energy, which results in similar profiles to that of W7 (Fig.~\ref{fig:density_comp}). Therefore while the composition of the ejecta differs between the W7- and N100-like models, our fits for SN~2011fe argue for a consistent density profile. 

\par

Unlike most other model parameters, we find significant differences in the location of the inner boundary velocity between our W7- and N100-like models. The location of the inner boundary velocity is highly model-dependent as this sets the amount and composition of material above the photosphere, which determines the overall spectrum. In addition, the inner boundary and composition, in combination with the time since explosion and luminosity, will also strongly impact the overall temperature of the model. It is therefore unsurprising that we generally find different inner boundary velocities for the W7- and N100-like models. Around maximum light our fits using both W7- and N100-like models find inner boundary velocities that are likely too low and indeed lie towards the lower limit of our training dataset, 4\,000~km~s$^{-1}$. Such velocities are significantly lower (by up to $\sim$3\,500~km~s$^{-1}$) than previous custom fitting of SN~2011fe (see Sect.~\ref{sect:comparisons}; \citealt{mazzali--14}). While this could point to a limitation in the construction of our training dataset, lower inner boundary velocities are instead likely a consequence of fitting the entire spectrum, with no weighting for features, and the photospheric approximation made by \textsc{tardis}. The result is that the fits will systematically favour low and potentially unrealistic inner boundary velocities, which produce spectra with higher temperatures that better match the overall shape of the observed spectra. This point is discussed further in Sect.~\ref{sect:weightings}.

\par

Overall, we find that \texttt{riddler} is able to produce reasonable fits to SN~2011fe using both W7- and N100-like ejecta structures. From \textsc{ultranest}, we find the evidence for our W7-like models is $\ln \mathcal{Z} \sim -538059$, while for our N100-like models $\ln \mathcal{Z} \sim -538366$. Calculating the Bayes factor (Eqn.~\ref{eqn:bf}), we find $\ln \mathrm{BF} = 307$. \cite{kaas--95} argue that a Bayes factor of $2\ln \mathrm{BF} >10$ indicates a very strong preference for a given model. Therefore our fitting indicates that a W7-like ejecta structure is heavily favoured, however we stress that this is dependent on our likelihood function, which should be treated cautiously when fitting full spectra, and our assumed priors, which do not cover the full parameter space required for N100-like models (Fig.~\ref{fig:11fe_params}). Considering individual spectra and their likelihoods however, the somewhat more mixed structure of N100 (compared to the highly stratified W7) is favoured at very early phases shortly after explosion.

\subsubsection{SN~2013dy}
\label{13dy}
We fit the HST spectra of SN~2013dy presented by \cite{pan--15}. We assume a distance modulus of $\mu = 30.68$ mag and total extinction of $E(B-V) = 0.35$ mag \citep{pan--15}. The best-fitting W7 and N100 spectra are shown in Fig.~\ref{fig:13dy_w7_all_mask}~\&~\ref{fig:13dy_n100_all_mask}, with posterior distributions shown in Fig.~\ref{fig:13dy_w7_all_mask_posterior}~\&~\ref{fig:13dy_n100_all_mask_posterior}, in the Appendix.

\par

In general, we find that SN~2013dy shows visually somewhat worse agreement with both W7- and N100-like models compared to SN~2011fe. At $-6.6$\,d both sets of models have similar likelihood ranges and show stronger spectral features and bluer spectra than SN~2013dy. As with SN~2011fe, we find our N100-like models favour lower kinetic energies, but we also find similarly low kinetic energies for our W7-like models. Given that N100 scaled to $1.0 \times 10^{51}$~erg produces a similar density structure to W7 (Fig.~\ref{fig:density_comp}), this would again imply that our N100 training dataset does not extend to sufficiently low energy to reproduce the overall density structure observed. Towards maximum light, both models show improved agreement with most spectral features, including \sistff\, and \cahk, but over-predict the NUV flux, as in the case of SN~2011fe. 

\par

With the exception of the first epoch, we find that W7-like models produce lower likelihoods compared to N100-like models. Again this is due to the better agreement between the model spectra and the continuum at longer wavelengths. At $-6.6$\,d, our neural networks show overlapping likelihoods for both models, indicating no strong preference for either. We also note that each of our N100 neural networks show considerably larger variations in their predictions than our W7 or SN~2011fe fits. This likely arises from the overall poor quality of the fits.

\par

From Fig.~\ref{fig:13dy_n100_all_mask_posterior} we find that N100-like models have significantly larger systematic differences relative to SN~2013dy than W7-like models ($\log f \sim30\%$ compared to $\sim$23\%), which is consistent with the generally worse quality of fits. As evident from the range of best-fitting spectra in Fig.~\ref{fig:13dy_n100_all_mask}, our N100-like models can show large variations in the best-fitting parameters. Across all neural networks we find much broader ranges of best-fitting values than for SN~2011fe, with NN 18 also showing a bi-modal distribution for the explosion epoch, kinetic energy, and inner boundary velocities. As was the case for SN~2011fe, some fits find inner boundary velocities that are too low and close to the lower limit of our training data, 4\,000~km~s$^{-1}$. This indicates that either our training dataset does not cover the full required range, or more likely the uniform weighting scheme is not appropriate at these phases.

\par

Overall, we find that \texttt{riddler} produces reasonable fits to SN~2013dy using W7-like models, but generally fails to find reasonable fits with an N100-like structure. Comparing the evidences calculated by \textsc{ultranest}, we find $\ln \mathrm{BF} = 1\,712$ -- indicating overwhelming evidence in favour of the W7 model across the full spectral sequence. Again we stress that the Bayes factor is useful only to determine the relative preference for a given model, based on our assumed priors and likelihood. It does not indicate that W7 is the correct model for a given SN and indeed our fitting indicates a wider parameter space is necessary for the N100 models but it is not clear if such a prior distribution would be physically realistic.

\subsection{Summary}

In summary, we applied \texttt{riddler} to fitting model spectra that were not seen by the neural networks during training. We showed that we are able to recover the input parameters of the models to within a few percent ($\lesssim$5\%). Applying \texttt{riddler} to the well-observed SNe~Ia, SN~2011fe and SN~2013dy, we find models containing a W7-like ejecta structure are generally preferred and able to reproduce observations within approximately one week of maximum light. For SN~2011fe, we find the earliest spectra (more than approximately 10\,d before maximum light) are better fit by models with an N100-like ejecta structure, while for SN~2013dy we find that N100-like ejecta structures produce significantly worse fits at all epochs. Based on the Bayes factor, we find strong evidence in favour of a W7-like model compared to an N100-like model when considering the full spectral sequence for both SN~2011fe and SN~2013dy. Again we stress that these conclusions are based on our assumed likelihood function and priors, and different models may be affected by the prior distributions in different ways.

%

\begin{figure*}
\centering
\includegraphics[width=\textwidth]{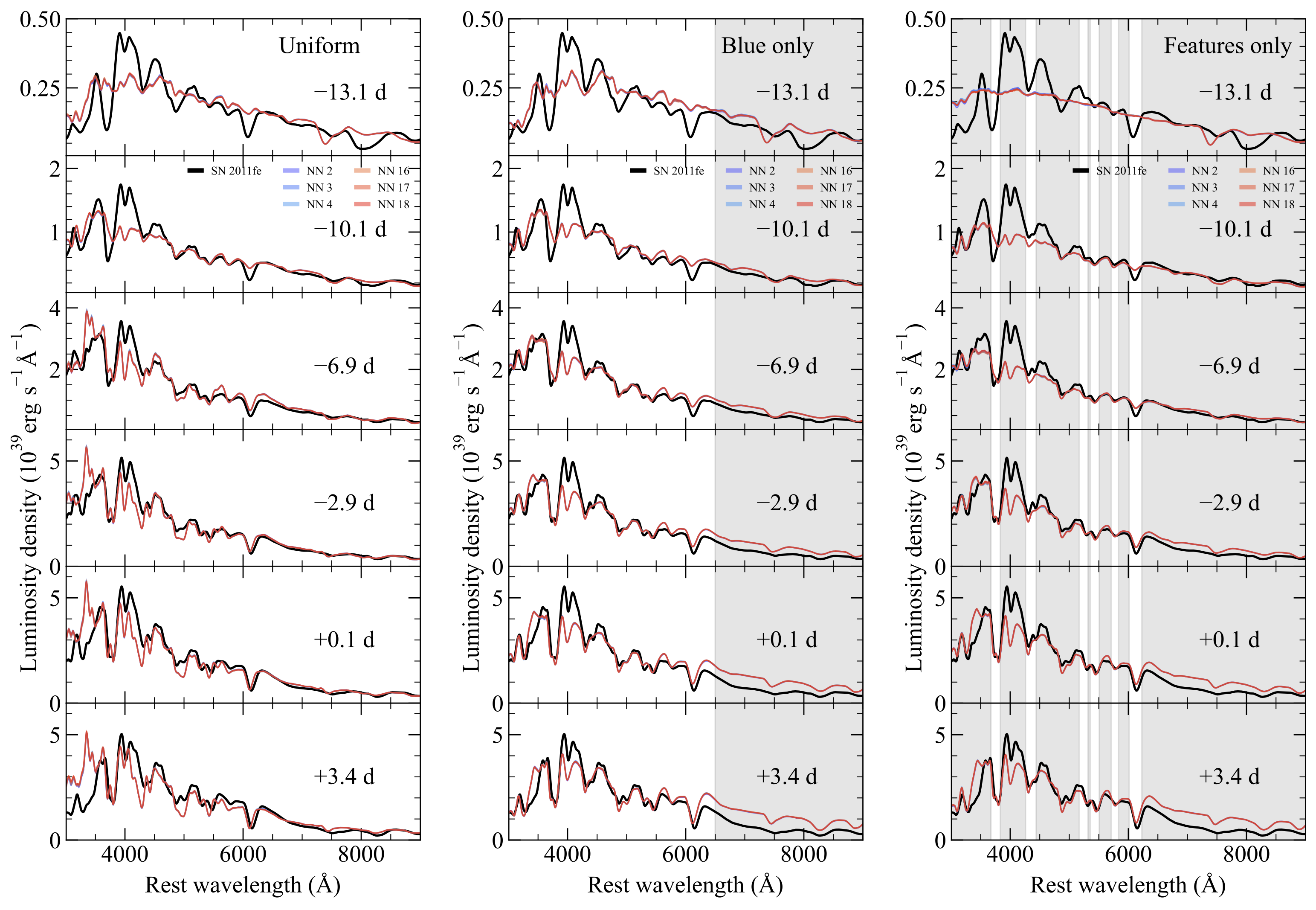}
\caption{Comparison between SN~2011fe and neural network reconstructions of the best-fitting W7 model spectra for each neural network and assuming different weighting schemes. Spectra are shown on an absolute luminosity scale with no additional scaling or offsets applied. Phases are given relative to $B$-band maximum. Grey shaded regions are not included during fitting.}
\label{fig:11fe_w7_all_mask}
\centering
\end{figure*}

\begin{figure*}
\centering
\includegraphics[width=\textwidth]{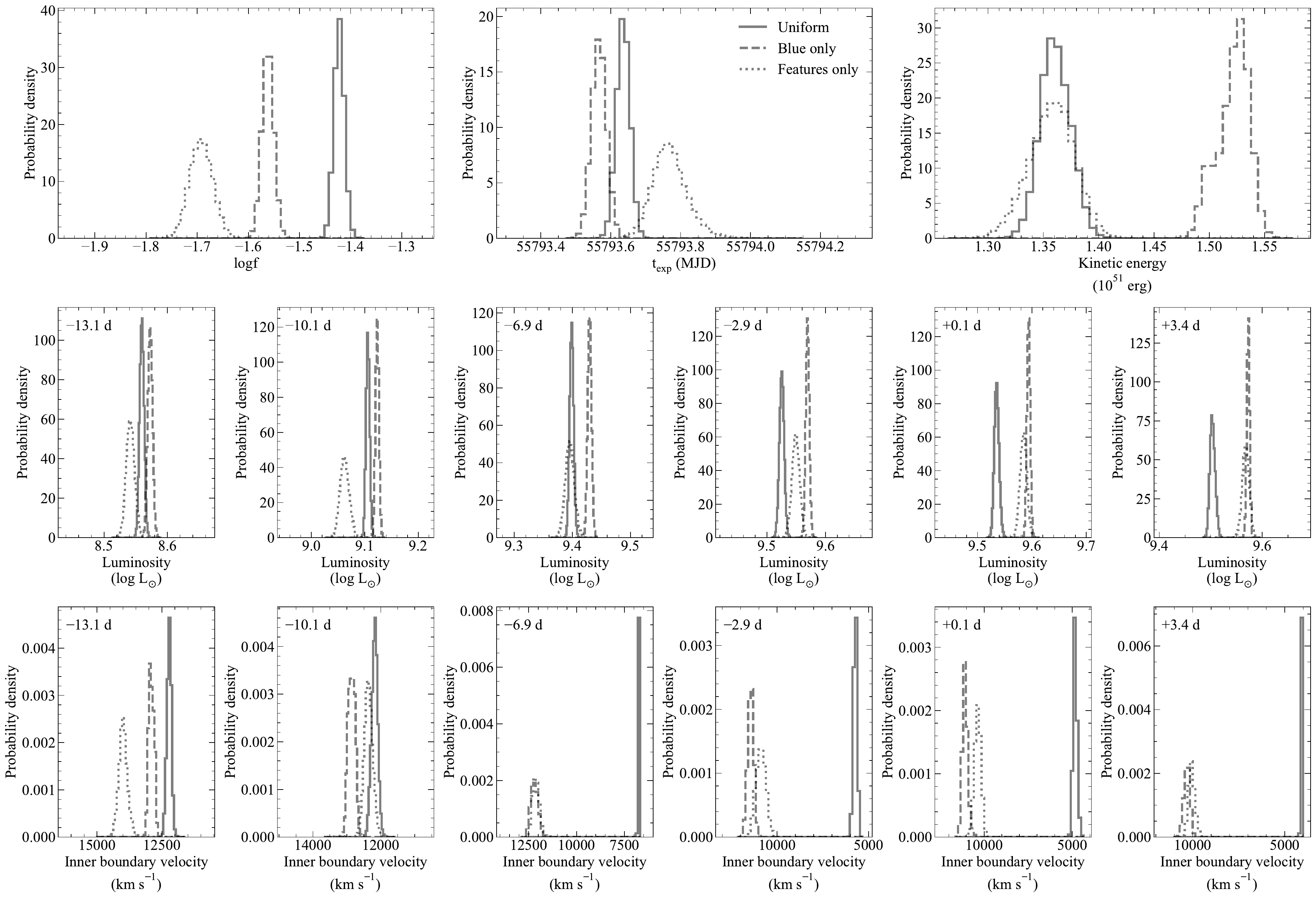}
\caption{Total posterior distributions for best-fitting parameters to SN~2011fe using W7 models across all neural networks fit and assuming different weighting schemes. Phases are given relative to $B$-band maximum.}
\label{fig:11fe_w7_all_mask_posterior}
\centering
\end{figure*}

\section{Impact of alternative weighting schemes}
\label{sect:weightings}

In Sect.~\ref{sect:observations}, we applied \texttt{riddler} to fit spectra of SNe~Ia from 3\,000 -- 9\,000~\AA. During our fitting process, we made the simplest assumption of treating all wavelengths uniformly with no additional weighting or flux scaling applied. The photospheric approximation used by \textsc{tardis} however means that certain wavelengths are expected to have larger systematic offsets than others and therefore should not necessarily be treated equally.

\par

\textsc{tardis} assumes a sharp boundary separating optically thick and thin regions. This allows \textsc{tardis} to model SNe spectra quickly by avoiding simulating regions of the ejecta with high optical depths. Due to this approximation, all Monte Carlo packets are injected into the simulation at the same physical position in the ejecta, regardless of their wavelengths. The optical depth at longer wavelengths however is generally lower than at shorter wavelengths, therefore the photosphere at longer wavelengths generally should be deeper inside the ejecta. By injecting all packets at the same location, those with longer wavelengths are typically able to escape the ejecta more easily than if they were injected deeper inside the ejecta. This usually manifests as an increased flux at longer wavelengths and is commonly observed closer to and beyond maximum light as the photospheric approximation becomes increasingly questionable (e.g. \citealt{stehle--05, mazzali--14}). Therefore, it is reasonable to allow for a systematically higher flux at longer wavelengths during our fitting process. 

\par

To demonstrate the impact of this, we perform additional fits in which wavelengths longer than $6\,500$~\AA\, were excluded, which we call `blue only' weighting. We note that the choice of $\lambda$~\textgreater~$6\,500$~\AA\, is somewhat arbitrary. The flux excess induced by the photospheric approximation becomes more pronounced at longer wavelengths, but does not necessarily begin at $6\,500$~\AA. This value was chosen simply to demonstrate its impact and due to the presence of few spectral features. We also test a third weighting scheme, `features only' weighting, in which only specific spectral features are included. This is designed to broadly mimic the approach taken by \cite{ogawa--23}, which was partially motivated by presence of weak-spectral features that are typically not well-fit by overall likelihood estimates (as is the case for \texttt{riddler}). We note however that we do not include additional comparisons between the model and data calculated by \cite{ogawa--23}, such as equivalent widths and velocity minima, which could yield better agreement in some cases (see Sect.~\ref{sect:observations}). For these tests and demonstrative purposes, the different schemes are applied at all epochs, but we again stress that \texttt{riddler} allows for any arbitrary weighting scheme, including those that vary with both wavelength and phase.

\par

Figure~\ref{fig:11fe_w7_all_mask} shows our best-fitting W7-like spectra compared to SN~2011fe assuming each of our three weighting schemes. Shaded regions are not included during the fits. The resulting posterior distributions are given in Fig.~\ref{fig:11fe_w7_all_mask_posterior}, while best-fitting parameter values are given in Table~\ref{tab:masks_spec_log_w7} in the Appendix. Comparisons for each weighting scheme against N100-like models are shown in the Appendix in Fig.~\ref{fig:11fe_n100_all_mask}, with posterior distributions shown in Fig.~\ref{fig:11fe_n100_all_mask_posterior} and best-fitting parameters given in Table~\ref{tab:masks_spec_log_n100}.  Similar figures for SN~2013dy are also given in Appendix~\ref{sect:appdx:figures_13dy}, again with best-fitting values in Tables~\ref{tab:masks_spec_log_w7}~\&~\ref{tab:masks_spec_log_n100}.

\par

From Fig.~\ref{fig:11fe_w7_all_mask}, it is clear that alternative weighting schemes, including only fitting specific spectral features, do not find improved agreement between SN~2011fe and W7-like models for the earliest spectra -- the outer ejecta of W7 simply does not contain the required structure to reproduce the spectral features observed. Beginning at $-6.9$\,d however we find model fits that do not include wavelengths $\lambda$~\textgreater~$6\,500$~\AA\, provide significantly better fits to spectral features, including the iron blends around $\sim$4\,500~\AA\, and $\sim$5\,000~\AA, and the NUV. Improved agreement with the NUV is a direct result of excluding longer wavelengths from the fit and therefore reducing the impact of the photospheric approximation. As longer wavelengths typically show an excess of flux, the best-fitting models will preferentially be those with higher temperatures, which naturally shifts more flux from longer to shorter wavelengths. This will better match the former at the expense of the latter, resulting in an excess of flux in the NUV. This is further demonstrated by the fact that Fig.~\ref{fig:11fe_w7_all_mask_posterior} shows these models also have systematically higher inner boundary velocities and therefore lower temperatures. For SN~2011fe, we find similar results for our feature weighting scheme (Fig.~\ref{fig:11fe_w7_all_mask}). Fits with N100-like models also show similar levels of improved agreement compared to SN~2011fe when using our alternative weighting schemes (Fig.~\ref{fig:11fe_n100_all_mask}).

\par

Considering the likelihood values, we find some changes in the best-fitting models depending on the weighting scheme. With uniform weighting, all spectra after $-10.1$\,d show higher likelihoods for W7-like models. Our blue only and features only weighting schemes however show slightly higher likelihoods for N100-like models between $-10.1$ -- $-2.9$\,d. Despite the majority of spectra showing higher likelihoods for N100-like models, \textsc{ultranest} estimates an overall higher evidence in favour of W7-like models, which is primarily driven by the much higher likelihood for the $+3.4$\,d spectrum. We note however that while the Bayes factor does still indicate a strong preference for the W7 model relative to N100, this is significantly reduced compared to uniform weighting. Across all neural networks, we find $\ln \textrm{BF} \sim$12 -- 60 for blue only weighting compared to $\sim$300 for uniform weighting. 

\par

For SN~2013dy, our uniform weighting scheme favours W7-like models for all epochs. With our blue weighting scheme, N100-like models are favoured for the first epoch only, however the difference in likelihoods is sufficiently large that the evidence is strongly in favour of an N100-like model overall ($\ln \mathrm{BF} \sim$9 -- 30). Indeed, our blue weighting scheme significantly reduces the scatter in best-fitting parameters for N100-like models and also produces visually better agreement, although we note that these models do somewhat over-predict the flux even at wavelengths \textless$6\,500$~\AA. This is again likely due to the photospheric approximation and the arbitrary wavelength cut-off used here. For our features weighting scheme, we again find W7-like models are favoured at most epochs, with N100-like being favoured only for the $-2.5$\,d spectrum. 

\par

In summary, we have shown how the best-fitting model parameters, and even ejecta structure, can be significantly impacted by the choice of weighting schemes used during fitting. Given the known limitations of the photospheric approximation we argue that, of the three simple schemes presented here, our blue weighting scheme provides the best compromise between fitting spectral features and continuum shape. More complicated weighting schemes are beyond the scope of this paper, but should be explored further. Such schemes include those that do not exclude longer wavelengths entirely, but instead allow for a systematically higher flux, and those that vary the weights as a function of phase, in addition to wavelength. Our results show that the best-fitting model parameters, and even the overall preferred explosion model, are sensitive to somewhat arbitrary choices of the likelihood function, demonstrated here by the omission of certain wavelengths. We stress that this quantitative fitting is nevertheless reproducible, due to the explicitly defined likelihood function and prior distributions. Manual fitting based on visual inspection is similarly arbitrary, but not reproducible due to the subjective nature with which model comparisons are made. We therefore strongly encourage future automated fitting routines and their applications to make explicitly clear what assumptions have been made during the fitting process and their expected impact.

%

\section{Discussion}
\label{sect:discussion}

Here we discuss the results of our \texttt{riddler} fitting. In Sect.~\ref{sect:limitations}, we discuss the limitations and assumptions of this work and where caution should be applied when using \texttt{riddler} to fit observations of SNe~Ia. In Sect.~\ref{sect:comparisons} we compare the results of our fitting to previous studies and alternative fitting methods. Finally, the quantitative method of spectral fitting outlined here allows for inferences about the SN explosion physics. In Sect.~\ref{sect:constraints} we discuss constraints on the explosion physics derived from our fitting.

\subsection{Limitations and assumptions}
\label{sect:limitations}

As discussed in Sect.~\ref{sect:weightings}, the assumed wavelength-dependent weighting used during fitting can have a significant impact on the best-fitting model parameters and therefore care must be taken when applying automated fitters to observations. Conclusions can only be drawn within the boundaries of the assumed goodness-of-fit metric and priors. Here we discuss other assumptions and limitations of this work.

\par

As with any machine learning-based technique, this work is limited by the datasets used to train our neural networks. We present neural networks trained on two Chandrasekhar mass explosion models, W7 and N100. Both models have been extensively studied within the literature and in general show good agreement with spectra of SNe~Ia, but some differences are clearly apparent (see e.g. \citealt{branch--85, sim--13}). We therefore expect that in general our neural networks will not be able to reproduce all features observed in SNe~Ia. Nevertheless, the relative level of agreement between different models can still provide useful insight into the explosion physics. While similar fractional offsets are typically not calculated for customised manual fitting, given the similar levels of agreement with observations we expect they have similar values.

\par

In addition to these limitations, neither model is capable of capturing the full diversity among SNe~Ia and therefore our fitting is currently limited to `normal' SNe~Ia \citep{benetti--05, branch--ia--groups}. By confining our training data to explosion model predictions however, we ensure that all of our models are physically consistent. Allowing individual elemental abundances to vary independently could result in statistically well-fitting, but physically unrealistic models. Such a scenario would make it difficult to provide meaningful constraints on the explosion mechanism for observed SNe~Ia. Future work will include an expanded set of explosion models and mechanisms as the basis of our training datasets, thereby enabling further robust and quantitative constraints to be placed on the explosion physics for a larger and more diverse sample of SNe~Ia.

\par

Aside from the underlying structure of the ejecta, when constructing our training datasets we also adopted a few additional assumptions. To determine a suitable range for the input luminosity and photospheric velocity, we used observations and measurements of SNe~Ia from \cite{scalzo--19} and \cite{foley--11} that were calculated relative to maximum light. Our \textsc{tardis} simulations however require a spectral phase relative to the time of explosion. We therefore assumed a rise time of 18.5\,d for all SNe~Ia. While some SNe~Ia do show rise times of $\sim$18.5\,d, significant diversity also exists, with rise times extending from as low as $\sim$15 up to $\sim$25\,d \citep{ganeshalingam--2011, firth--sneia--rise, miller--20a}. The exact measured value of the rise time is dependent on both the model assumptions and the data quality used during fitting \citep{miller--20a}. Future work will include a wider range of rise times within the training datasets, which will likely be necessary as other explosion models with different ejecta masses are included \citep{scalzo--14}. We note however that given the wide range of luminosities and velocities used in our training datasets, our fitting technique does not assume a rise time of 18.5\,d when applied to observations. This is demonstrated from our fits to both SN~2011fe and SN~2013dy, which imply rises times of $\sim$20.6\,d and $\sim$16 -- 20\,d, respectively (depending on the model and assumed weighting). We also set an arbitrary lower limit on the inner boundary velocity of our models of 4\,000~km~s$^{-1}$. As discussed in Sect.~\ref{sect:observations}, some of our best-fitting models favour inner boundaries at this lower limit, indicating our training dataset does not cover the full parameter space required to match the observations. Again however this only affects fits with uniform weighting of all wavelengths and arises from the photospheric approximation. Alternative schemes that account for this approximation do not show similarly skewed velocities. 

\par

The spectra used to train our neural networks were generated from 3\,000 -- 9\,000~\AA\ and therefore, our neural networks cannot currently be used to fit UV spectra. These wavelengths are highly sensitive to metallicity effects and can provide important constraints on the progenitor metallicity \citep{foley--kirshner--13, brown--15}. Future work will include an expanded wavelength range, but we again stress the importance of appropriate weighting schemes, which can significantly impact the results.

\par

Finally, during our fitting process we also set a prior constraint that the inner boundary velocity is always decreasing with time, which helps to improve the convergence of the \textsc{ultranest} fits. This has also generally been found by manual spectral fitting (e.g. \citealt{stehle--05, mazzali--14}). To test the impact of this explicit assumption, we run additional fits for both SN~2011fe and SN~2013dy with no prior constraint. For SN~2011fe, we find no significant differences for the best-fitting parameters, which do naturally show a decreasing inner boundary velocity with time. One exception however is the $+3.4$\,d spectrum for which we find a best-fitting inner boundary velocity closer to the previous $+0.1$\,d spectrum ($\sim$4\,840~km~s$^{-1}$) rather than the lower limit of our training dataset (4\,000~km~s$^{-1}$). Both fits however are also consistent with each other within the full uncertainty range. In the case of SN~2013dy, we find noticeable changes in the best-fitting parameters and that the inner boundary velocity does not monotonically decrease with time. Again we note that the N100-like models show overall poor agreement with SN~2013dy when considering a uniform wavelength weighting. Assuming our blue weighting scheme however, which shows significantly better agreement, our best-fitting models recover a monotonically decreasing inner boundary velocity and predict comparable best-fitting parameters to those with a prior velocity constraint. We therefore argue that the assumed prior constraint on the velocity evolution is unlikely to have a significant impact on models that provide a good match to the observations. For well-sampled spectral sequences, with much smaller time intervals between subsequent observations, a somewhat relaxed velocity constraint may be necessary.

\subsection{Comparisons with existing models and methods}
\label{sect:comparisons}
The spectra of SN~2011fe fit with \texttt{riddler} in Sect.~\ref{sect:observations} were the subject of a previous analysis by \cite{mazzali--14}. \cite{heringer--17} also present spectral comparisons between these spectra and \textsc{tardis} models calculated using similar input parameters. Here we compare the results of our fits with these previous results. We note that model spectra presented by \cite{mazzali--14} were calculated with a different radiative transfer code and therefore we expect some discrepancies between best-fitting parameters and spectra to arise due to differences in the radiative transfer treatments and atomic data \citep{heringer--17}. \cite{mazzali--14} present spectra calculated for SN~2011fe using both the W7 and WS15DD1 models \citep{iwamoto--99}. The WS15DD1 model is a delayed detonation explosion similar to the N100 model used here however we note that the WS15DD1 model was calculated in one dimension, has a slightly smaller $^{56}$Ni mass, and shows a different abundance distribution. 

\par

Based on their spectral fits with the W7 model, \cite{mazzali--14} argue for an explosion epoch of MJD = 55\,795.2$\pm$0.5, whereas our \texttt{riddler} fits predict an earlier explosion epoch (by $\sim$1.6$\pm$0.7 d) of MJD = 55\,793.63$\pm$0.5 (across all of the weighting schemes included here). \cite{mazzali--14} show how differences in the rise time can have a significant impact on the earliest spectrum at $-15.3$\,d, with longer rise times (i.e. earlier explosion epochs) producing worse agreement. Spectra even $\sim$2\,d later however are less sensitive to the rise time and show only minor variations. The $-15.3$\,d spectrum modelled by \cite{mazzali--14} was not included in our fits as the estimated phase relative to explosion was outside the time range covered by our input parameters (5 -- 25\,d). This likely explains the difference in rise time estimates and further highlights the need for spectra as early as possible to place tight constraints on the explosion epoch through spectral modelling. We note that both explosion epoch estimates are still earlier than those predicted from the light curve \citep{11fe--nature, magee--20}.

\par

Overall, we find qualitatively similar levels of agreement between our best-fitting spectra and those presented by \cite{mazzali--14} (and \cite{heringer--17}) for our blue only weighting scheme, which we consider a more appropriate comparison. Both sets of models generally predict similar velocities and widths for most spectral features. We find that our spectra typically provide better agreement with the relative strengths of features, such as the \ion{S}{ii} W feature and the iron blends around $\sim$4\,500 and 5\,000~\AA, however the \cite{mazzali--14} spectra provide better matches to the NUV. Comparing the model input parameters, we find that our luminosities are comparable to those estimated by \cite{mazzali--14} and differ by $\lesssim$0.1~dex. The inner boundary velocities however show significant differences, with our values being systematically higher (by up to $\sim$3\,900~km~s$^{-1}$) than those of \cite{mazzali--14} when assuming a blue only weighting scheme. For our uniform weighting scheme, we find our inner boundary velocities are instead systematically lower (by up to $\sim$3\,500~km~s$^{-1}$). We note that the inner boundary velocities presented by \cite{heringer--17} also show some differences (50 -- 850~km~s$^{-1}$) relative to \cite{mazzali--14}, despite the same explosion epoch and luminosity. Given that many of our spectral line ratios provide better matches to SN~2011fe when assuming a blue only weighting scheme, this would indicate that the temperature and ionisation state of these models is closer to that of SN~2011fe. This further demonstrates the importance of an appropriate weighting scheme to account for excess flux due to the photospheric approximation. 

\par

Aside from spectra calculated with existing explosion models, by modelling progressively later spectra \cite{mazzali--14} also develop a custom ejecta profile. This custom profile is motivated primarily by differences in the UV spectra and provides better qualitative agreement with SN~2011fe spectra. This profile contains a tail of material extending to higher velocities than in the W7 model, but with densities lower than the WS15DD1 delayed detonation model. Similarly, \cite{mazzali--14} find a more mixed composition for the ejecta produces better agreement than the standard W7 model. Although we do not construct custom ejecta models and do not consider the UV here, we find qualitatively similar results. Both our best-fitting W7- and N100-like models predict similar density profiles. With our approach however, neither best-fitting model has the freedom to extend the density profile to higher velocities. In addition, we find that the less stratified structure of the N100-like models produces better agreement with many of the spectral features in SN~2011fe, particularly at early times. Therefore, even though our models do not have complete freedom when fitting individual SNe, comparative analysis between multiple explosion models can provide useful insights without the need for customisation of models for each SN.

\par

Finally, we also compare our results to a \textsc{dalek} fit of the $-10.1$\,d SN~2011fe spectrum using the method outlined by \cite{obrien--23}. This spectrum was selected by \cite{obrien--23} as it lies approximately 8 -- 12\,d post-explosion. We note that this \textsc{dalek} fit is performed using continuum normalised spectra and therefore we cannot directly compare on an absolute flux scale, as in our \texttt{riddler} fits. Based on our \textsc{dalek} fit, we find an explosion epoch of 55\,794.3$\pm$0.2, which lies approximately between the values predicted by our fits and \cite{mazzali--14}. We also find an inner boundary velocity of $\sim10\,800\pm160$~km~s$^{-1}$, which is lower than the best-fitting value for our blue only weighted N100-like model (the best match at this epoch, $\sim$11\,900 -- 13\,300~km~s$^{-1}$) and marginally lower than the \cite{mazzali--14} custom model (11\,300~km~s$^{-1}$). 

\par

Using the masses determined by our \textsc{dalek} fit for the $-10.1$\,d SN~2011fe spectrum, we use equations 1 -- 7 in \cite{obrien--23} to reconstruct the best-fitting ejecta structure. We note however that the simulation only includes material above the inner boundary ($10\,800\pm160$~km~s$^{-1}$ at this epoch). We find that the structure of the \textsc{dalek} model shows a generally shallower density slope and lower overall density than the W7 or N100 models. This lower overall density is the likely cause of the lower inner boundary velocity, as the photosphere needs to be placed deeper inside the model to produce the required temperature and features. Indeed, the density at the inner boundary is comparable across all of the models. The abundance distributions also show some differences. From our \textsc{dalek} fit, the distribution of intermediate mass elements peaks at $\sim$14\,000~km~s$^{-1}$. This is comparable to the N100 model, but slightly higher than our best-fitting W7- and N100-like models ($\sim$11\,000 -- 12\,000~km~s$^{-1}$). The velocity range of intermediate mass elements is also comparable to the W7 model, although more narrow than N100. At the inner regions of the ejecta model, our \textsc{dalek} fit is dominated by $^{56}$Ni, while both our W7- and N100-like models show more mixed structures with some $^{56}$Ni, other iron-group elements, and a significant fraction of intermediate mass elements. Despite the different training data and methods used, we find qualitatively consistent results between our \texttt{riddler} and \textsc{dalek} fits, although again we note that we are unable to directly compare the absolute luminosities. Both methods indicate a larger fraction of intermediate mass elements at high velocities relative to the standard W7 model is required to reproduce the early spectra of SN~2011fe.

\par

In summary, we find that our \texttt{riddler} fits are able to produce comparable best-fitting spectra to existing methods using either customised or automated fitting. Current techniques rely on modelling of progressively later spectra, as the photosphere recedes deeper inside the ejecta, to determine customised abundance profiles. As new layers of the ejecta are exposed, the composition in the outer regions is held fixed. This ensures a self-consistent model. A key benefit of \texttt{riddler} however is that all spectra are fit simultaneously, naturally ensuring a consistent ejecta profile, and reproducing the luminosity and temperature evolution of the observed SN.

\subsection{Explosion physics constraints}
\label{sect:constraints}
Our fits to SN~2011fe and SN~2013dy show that neither the W7 nor N100 explosion model is able to reproduce all features of either SN, however different regions of the ejecta do show better agreement with individual models. At early times, we find that the N100 model produces better agreement with SN~2011fe than W7, while at later times this is reversed. Such changes in the best-fitting models at different phases can provide further constraints on the explosion physics.

\par

The outer regions of the W7 model are dominated by unburned carbon and oxygen, following quenching of the deflagration front \citep{nomoto--84}. Consistent with previous studies, our results show that these outer layers in the W7 model cannot reproduce early observations of SNe without some additional mixing \citep{branch--85, stehle--05}. Relative to SN~2011fe, our W7-like models show stronger carbon and oxygen features, in particular \ion{C}{ii}~$\lambda$6\,578 and \ion{O}{i}~$\lambda$7\,774,
and significantly weaker intermediate mass element features at these times. While features indicative of unburned material have been detected in SN~2011fe and other SNe~Ia at early times \citep{11fe--nature, parrent--2012, folatelli--12}, they are not as strong as those shown by our models. This would indicate that enhanced burning is required or increased mixing to reduce the abundance in the outer layers and extend this distribution down to lower velocities. For the N100 model, unburned fuel is also found in the outer ejecta, but these regions are instead mostly dominated by a larger fraction of burned material in the form of intermediate mass elements \citep{seitenzahl--13}. Unburned material does however extend to much lower velocities than in the W7 model. The \ion{O}{i}~$\lambda$7\,774 feature predicted by our N100 models shows good agreement with the shape of the feature around $\sim$7\,500~\AA\, in SN~2011fe, but is slightly stronger than observed for the earliest spectrum at $-13.1$\,d. This could indicate that while the overall velocity-distribution is similar to SN~2011fe, a somewhat reduced mass fraction is preferred for the outermost ejecta at least. Such differences in the distribution of burned and unburned material are a natural consequence of buoyancy \citep{khokhlov--95, townsley--07}. In multi-dimensional simulations of deflagrations, buoyancy of the deflagration ash drives plumes towards the outer ejecta, while in one-dimensional models, such as W7, no such buoyancy can occur \citep{pakmor--24}. This leads to a highly stratified ejecta with burned material from the high density regions of the white dwarf confined to the inner ejecta. 

\par

None of our best fitting models are able to reproduce the high-velocity \ion{Ca}{ii}~NIR features observed in SN~2011fe. High-velocity features are commonly observed in spectra of SNe~Ia up to maximum light at velocities $\sim$5\,000 -- 10\,000~km~s$^{-1}$ above the photosphere \citep{mazzali--05, maguire--14, silverman--15}. Although the cause of these features is currently unknown, it has been suggested that they could arise from properties of the progenitor system, through interaction with circumstellar material, or from properties of the explosion, as certain explosion scenarios predict high-velocity shells of material from incomplete silicon burning \citep{wang--03, gerardy--04, mazzali--05}. \cite{magee--21} present models of double detonation explosions \citep{bildsten--07, kromer--10, shen--14, polin--19} and demonstrate the impact of a high-velocity shell of material on the \sistff\,
feature at early times. In addition, \cite{clark--21} recently argued against circumstellar interaction based on multiple probes believed to be linked to circumstellar interaction and the lack of any correlation. The W7 and N100 models implemented into the current version of \texttt{riddler} do not contain high-velocity shells of material or circumstellar interaction, therefore the lack of agreement with the high-velocity features is unsurprising. Nevertheless our models do cover a wide range of ionisation states for ejecta. The fact that these are unable to simultaneously match the photospheric and high-velocity features indicates additional complexity is required. Future versions of \texttt{riddler} will incorporate models that do contain high-velocity shells and therefore may be able to provide direct evidence in favour of or refute specific interpretations.

%

\section{Conclusions}
\label{sect:conclusions}

In this work, we presented \texttt{riddler}, a method for automated and quantitative fitting of SNe~Ia spectral time series, beginning a few days after explosion up to shortly after maximum light. Using predictions from the well-studied W7 \citep{nomoto--84} and N100 \citep{seitenzahl--13} explosion models, we developed datasets consisting of 100\,000 spectra per model with the \textsc{tardis} radiative transfer code \citep{tardis}. These datasets were then used to train a series of neural networks that act as emulators of full \textsc{tardis} simulations. Using our emulators, we fit spectra of SNe~Ia with \textsc{ultranest} \citep{ultranest} to quantify the relative likelihoods of explosion models and determine the best-fitting parameters and posterior distributions. Compared to previous studies using similar emulators, our models incorporate densities and compositions from realistic explosion models. They can therefore be used to directly quantify the relative agreement between multiple theoretically predicted ejecta structures. Our fits are also performed to multiple spectra of a given SN at different epochs simultaneously, thereby naturally producing a self-consistent model and matching the luminosity and temperature evolution of the observed SN.

\par

To demonstrate the viability of \texttt{riddler}, we fit model spectra generated in the same manner as our training datasets, but not seen during training. We showed that \texttt{riddler} is able to recover the input parameters of the model with reasonable accuracy. We then used \texttt{riddler} to fit observations of two well-studied SNe~Ia, SN~2011fe \citep{11fe--nature} and SN~2013dy \citep{zheng--13}. Based on our fits, we showed that in general the W7 model is strongly favoured overall for our assumed likelihood and priors. Through qualitative comparisons and by considering the likelihoods of the models at each phase however, we showed that the earliest spectra are not well reproduced by the W7 model. Instead a more mixed ejecta (similar to the N100 model) is favoured, consistent with previous studies. Comparing our results to existing models and methods, we find comparable agreement between spectra.

\par

Using \texttt{riddler}, we also demonstrated the importance of different weighting schemes when quantifying the goodness-of-fit between model spectra and observations. Depending on the relative weighting of wavelengths or specific features, the best-fitting model parameters and spectra can show significant variation. This also applies in general to the likelihood function and prior distributions assumed when performing such fits, both of which can have a significant impact on the overall best-fitting model and parameters. Manual fitting of spectra however is typically performed in a qualitative and subjective manner, and is therefore not reproducible. Although the exact goodness-of-fit metric and priors do  require somewhat arbitrary choices to be made when using automated fitters, these choices can at least be made explicit and quantified, enabling reproducible and consistent studies of SNe spectra.

\par

In the coming years, spectral modelling of SNe~Ia will necessarily become increasingly automated. Manual fitting of individual spectra is both time and computationally intensive, and cannot provide a quantitative measurement of the relative agreement between models nor the uncertainties for their best-fitting parameters. The number of SNe~Ia spectra currently archived already dwarfs our ability to perform manual fitting of them. With large scale spectroscopic surveys, such as the SED Machine as part of Zwicky Transient Facility \citep{sedm, bellm--19} and TiDES on 4MOST \citep{tides}, this will continue to grow even larger. Automated fitters are therefore required to make continued progress in constraining the explosion physics of SNe~Ia. With \texttt{riddler}, we have developed a tool that enables entire spectral sequences to be fit simultaneously and allows for robust comparisons across explosion models. Future work will continue to improve the training dataset and neural networks used in \texttt{riddler}, including expansion to a wider variety of explosion models.

%

\section*{Acknowledgements}

We thank the referee for useful and constructive feedback that improved the clarity of our manuscript. 
MRM wishes to thank the organisers of the DEVISE (Decoding, Enabling, and deVeloping AI Tools for Research) AI/ML workshop and J. O'Brien for useful discussions. MRM acknowledges a Warwick Astrophysics prize post-doctoral fellowship made possible thanks to a generous philanthropic donation. KM is funded by the EU H2020 ERC grant no. 758638. Computing facilities were provided by the Scientific Computing Research Technology Platform of the University of Warwick, the Queen's University Belfast HPC Kelvin cluster, and by TCHPC (Research IT, Trinity College Dublin). This research made use of \textsc{Tardis}, a community-developed software package for spectral synthesis in supernovae \citep{tardis}. The development of \textsc{Tardis} received support from the Google Summer of Code initiative and from ESA's Summer of Code in Space program. \textsc{Tardis} makes extensive use of Astropy and PyNE. We derive posterior probability distributions and the Bayesian
evidence with the nested sampling Monte Carlo algorithm
MLFriends (Buchner, 2014; 2019) using the
UltraNest\footnote{\url{https://johannesbuchner.github.io/UltraNest/}} package (Buchner 2021). This work made use of the Heidelberg Supernova Model Archive (HESMA; \citealt{hesma}), https://hesma.h-its.org.

\section*{Data Availability}

\texttt{riddler} is publicly available on GitHub\footnote{\href{https://github.com/MarkMageeAstro/riddler}{https://github.com/MarkMageeAstro/riddler}}.



\bibliographystyle{mnras}
\bibliography{Magee}




\appendix

\section{Neural network training}
\label{sect:appdx:nn_accuracy}

\begin{table*}
\centering
\caption{Hyperparameters and performance for neural networks trained on W7 and N100 models}\tabularnewline
\label{tab:nn_hyper}\tabularnewline
\begin{tabular}{lllllcccc}\hline
\hline
\tabularnewline[-0.25cm]
    &  &   & &  &  \multicolumn{2}{c}{W7} & \multicolumn{2}{c}{N100}    \tabularnewline
Neural network      & Batch size & Learning rate        & Neurons  &  Layers   & Median Mean FE    & Median Max FE  & Median Mean FE    & Median Max FE  \tabularnewline
\hline
\multicolumn{9}{c}{Varying layers} \tabularnewline
\hline
1                    & 10\,000              & $1\times10^{-3}$              & 400               &  1            & 0.0166   	 		&  0.0712    & 0.0192 & 0.0852		\tabularnewline 
2                    & 10\,000              & $1\times10^{-3}$              & 400               &  2            & 0.0156 	 		&  0.0684    & 0.0174 & 0.0802		\tabularnewline 
3                    & 10\,000              & $1\times10^{-3}$              & 400               &  3            & 0.0155     		&  0.0679    & 0.0174 & 0.0802		\tabularnewline 
\textbf{4}           & \textbf{10\,000}     & \textbf{$1\times10^{-3}$}     & \textbf{400}      &  \textbf{4}   & \textbf{0.0156}   &  \textbf{0.0683}  & \textbf{0.0174} & \textbf{0.0801} \tabularnewline 
5                    & 10\,000              & $1\times10^{-3}$              & 400               &  5            & 0.0157     		&  0.0688    & 0.0174 & 0.0806		\tabularnewline 
6                    & 10\,000              & $1\times10^{-3}$              & 400               &  6            & 0.0158     		&  0.0693    & 0.0178 & 0.0811		\tabularnewline 
\hline
\multicolumn{9}{c}{Varying neurons} \tabularnewline
\hline
7                    & 10\,000              & $1\times10^{-3}$              & 100               &  4            & 0.0158     		&  0.0693    	& 0.0177 & 0.0809	\tabularnewline 
8                    & 10\,000              & $1\times10^{-3}$              & 200               &  4            & 0.0156     		&  0.0681    	& 0.0174 & 0.0799	\tabularnewline 
9                    & 10\,000              & $1\times10^{-3}$              & 300               &  4            & 0.0156     		&  0.0682    	& 0.0174 & 0.0802	\tabularnewline 
\textbf{4}           & \textbf{10\,000}     & \textbf{$1\times10^{-3}$}     & \textbf{400}      &  \textbf{4}   & \textbf{0.0156}   &  \textbf{0.0683}  & \textbf{0.0174} & \textbf{0.0801} \tabularnewline 
10                   & 10\,000              & $1\times10^{-3}$              & 500               &  4            & 0.0156     		&  0.0680    	& 0.0175 & 0.0800	\tabularnewline 
11                   & 10\,000              & $1\times10^{-3}$              & 600               &  4            & 0.0157     		&  0.0685    	& 0.0176 & 0.0805	\tabularnewline 
\hline
\multicolumn{9}{c}{Varying batch size} \tabularnewline
\hline
12                   & 20\,000              & $1\times10^{-3}$              & 400               &  4            & 0.0157     		&  0.0688    	& 0.0175 & 0.0799	\tabularnewline 
\textbf{4}           & \textbf{10\,000}     & \textbf{$1\times10^{-3}$}     & \textbf{400}      &  \textbf{4}   & \textbf{0.0156}   &  \textbf{0.0683}  & \textbf{0.0174} & \textbf{0.0801} \tabularnewline 
13                   & \phn{}5\,000         & $1\times10^{-3}$              & 400               &  4            & 0.0156     		&  0.0687    	& 0.0174 & 0.0797	\tabularnewline 
14                   & \phn{}2\,500         & $1\times10^{-3}$              & 400               &  4            & 0.0156     		&  0.0683    	& 0.0174 & 0.0800	\tabularnewline 
15                   & \phn{}1\,000         & $1\times10^{-3}$              & 400               &  4            & 0.0156     		&  0.0680    	& 0.0175 & 0.0799	\tabularnewline 
\hline
\multicolumn{9}{c}{Varying learning rate} \tabularnewline
\hline
\textbf{4}           & \textbf{10\,000}     & \textbf{$1\times10^{-3}$}     & \textbf{400}      &  \textbf{4}   & \textbf{0.0156}   &  \textbf{0.0683}  & \textbf{0.0174} & \textbf{0.0801} \tabularnewline 
16                   & 10\,000              & $5\times10^{-4}$              & 400               &  4            & 0.0155     		&  0.0682    	& 0.0173 & 0.0797	\tabularnewline 
17                   & 10\,000              & $1\times10^{-4}$              & 400               &  4            & 0.0155     		&  0.0681    	& 0.0173 & 0.0799	\tabularnewline 
18                   & 10\,000              & $5\times10^{-5}$              & 400               &  4            & 0.0155     		&  0.0681    	& 0.0173 & 0.0799	\tabularnewline 
19                   & 10\,000              & $1\times10^{-5}$              & 400               &  4            & 0.0156     		&  0.0688    	& 0.0174 & 0.0806	\tabularnewline 
20                   & 10\,000              & $5\times10^{-6}$              & 400               &  4            & 0.0157     		&  0.0692    	& 0.0175 & 0.0809	\tabularnewline 
\hline
\hline
\end{tabular}
\end{table*}

\begin{figure}
\centering
\includegraphics[width=\columnwidth]{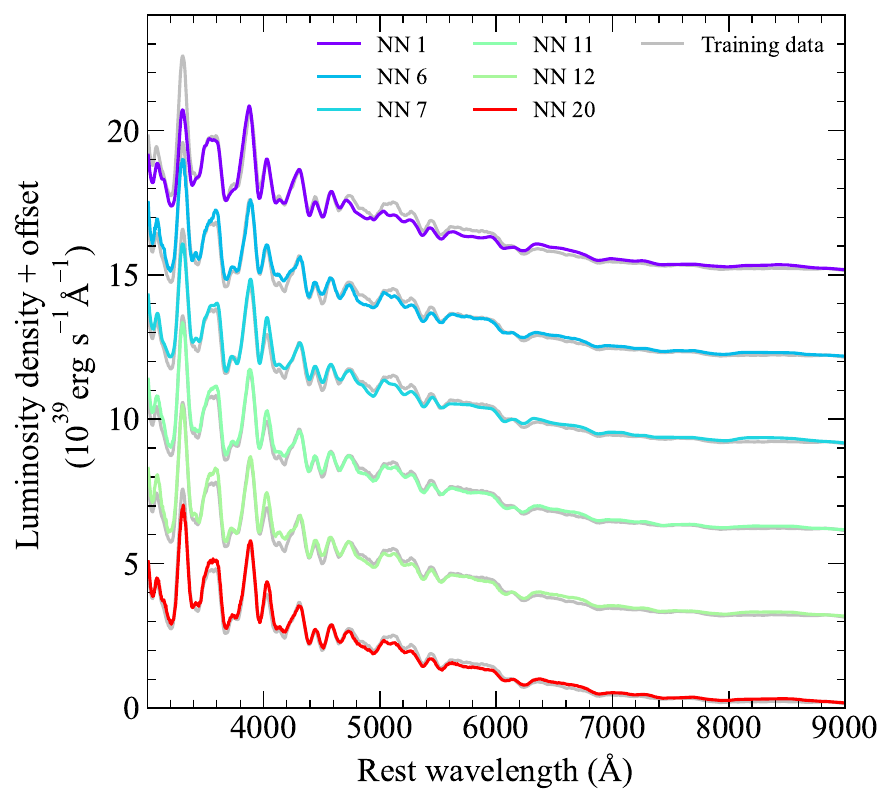}
\caption{Predicted spectra for our worst performing neural networks compared against the true model spectrum with the largest mean fractional error (grey). Predicted spectra show typical mean fractional errors of $\sim$9 -- 14\%. Maximum fractional errors for each predicted spectrum range from $\sim$40 -- 70\% and always occur between $\sim$8\,100~--~8\,300~\AA.}
\label{fig:nn_preds_high}
\centering
\end{figure}

\begin{figure}
\centering
\includegraphics[width=\columnwidth]{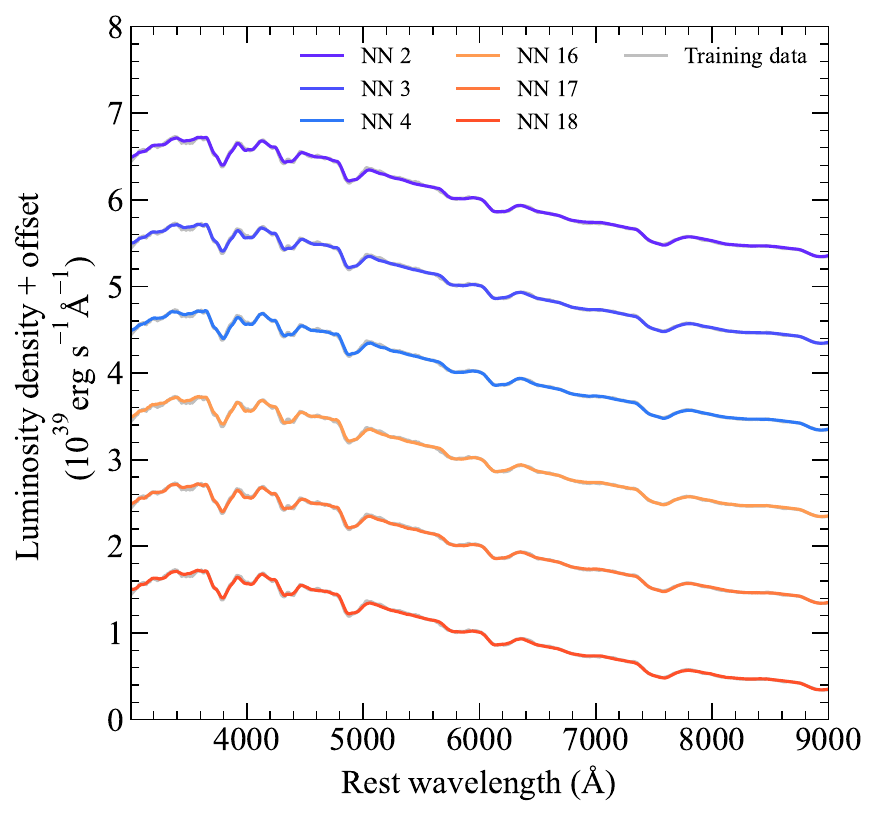}
\caption{Predicted spectra for our best performing neural networks compared against the true model spectrum with the smallest mean fractional error (grey). Predicted spectra show typical mean fractional errors of $\sim$0.8 -- 0.9\%. Maximum fractional errors for each predicted spectrum range from $\sim$3 -- 4\% and occur in either the first or last wavelength bin, with the exception of NN~16 for which the maximum error occurs at $\sim$8\,600~\AA.}
\label{fig:nn_preds_low}
\centering
\end{figure}

Here we provide a more detailed discussion of the accuracies for the neural networks used in this work. As discussed in Sect.~\ref{sect:nn}, to determine the accuracy of each neural network, we use our final set of 20\,000 testing models. These models were not previously seen during training of any of the neural networks. In Fig.~\ref{fig:nn_preds_high} we show predicted W7 spectra with the highest mean fractional errors from our worst performing neural networks, while in Fig.~\ref{fig:nn_preds_low} we show predicted spectra with the lowest mean fractional errors from our best performing neural networks. In general, the neural networks are able to reproduce most of the prominent features in the spectra, with the correct velocity and width, but struggle to reproduce weaker and more narrow features -- likely due to their overall lower contribution to the neural network loss during training. For our worst performing neural networks, Fig.~\ref{fig:nn_preds_high} shows that they struggle to reproduce the flux level for some spectral features and could be discrepant by $\gtrsim$10\%. For our W7 neural networks, the largest errors occur at longer wavelengths where the spectra do not show any strong features. Conversely, our N100 neural networks show the largest errors, potentially up to $\sim$20\% in extreme cases, at wavelengths $\lesssim$4\,000~\AA.

\par

\begin{figure}
\centering
\includegraphics[width=\columnwidth]{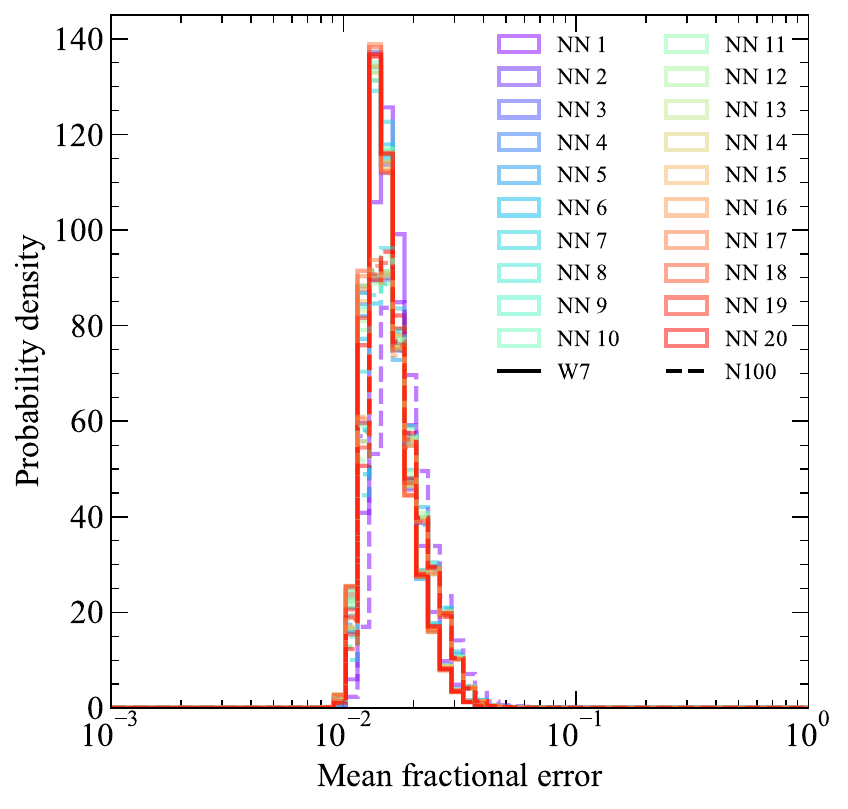}
\caption{Distribution of mean fractional errors across predicted spectra for the 20\,000 models in both our W7 and N100 testing datasets. }
\label{fig:nn_mfe_dist}
\centering
\end{figure}


\par

\begin{figure}
\centering
\includegraphics[width=\columnwidth]{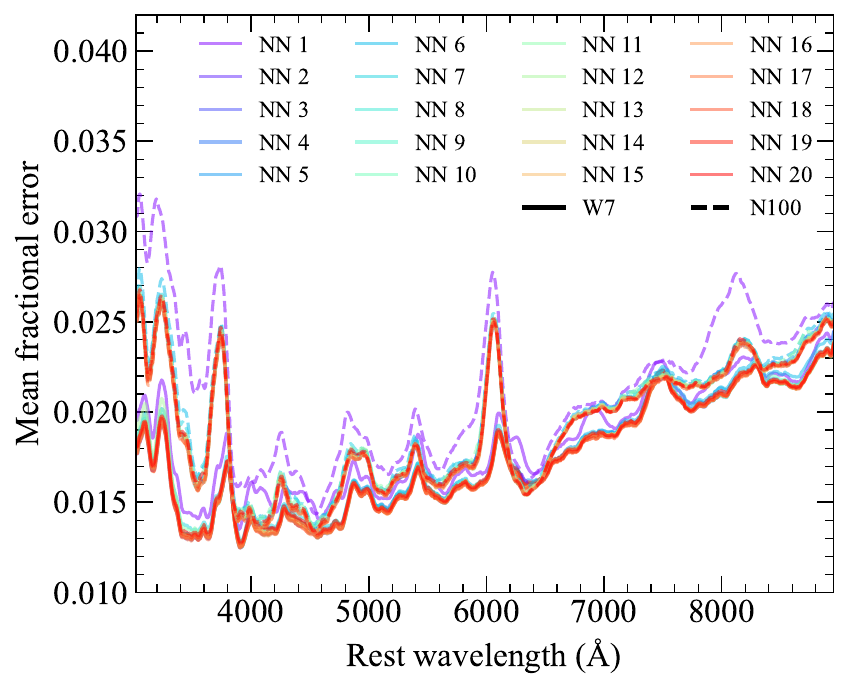}
\caption{Mean fractional error as a function of wavelength for the 20\,000 models in both our W7 and N100 testing datasets. }
\label{fig:nn_mfe_spec}
\centering
\end{figure}

In Fig.~\ref{fig:nn_mfe_dist}, we show the distribution of mean fractional errors across our 20\,000 W7 and N100 testing models. For both the W7 and N100 models, we find that most spectra show typical mean fractional errors of $\sim$1.6 -- 1.8\%, with the largest mean fractional errors being $\sim$5 -- 20\% across all of the neural networks. Unsurprisingly, we find the mean fractional errors of our predicted spectra are highly correlated with sparsely-populated regions of the training dataset parameter space. In particular, those models with velocities towards the lower limit for a given time typically show higher mean fractional errors. Figure~\ref{fig:nn_mfe_spec} shows the mean fractional error as a function of wavelength across the 20\,000 W7 and N100 testing models, which is generally highest in NUV and NIR spectral regions. To account for differences in the level of accuracy as a function of the model input parameters, we calculate the mean fractional error as a function of wavelength in a series of time and velocity bins. We then define an interpolator to calculate the mean fractional error for any values of time and velocity. When fitting spectra, we include this systematic offset, based on the input parameters of the each model spectrum, as an estimate of the neural network prediction error ($\Delta_{\lambda}(t, v)$; Eqn.~\ref{eqn:uncertainty}).

\par

In summary, we find our neural networks reach typical accuracies of a few percent and are likely limited by sparse regions of the input parameter space for our training dataset. The inclusion of active learning could aid in training the neural networks in these regions and will be explored in future work \citep{obrien--23}. In addition, future work will also explore larger training datasets and different methods of data augmentation, in addition to more complicated neural network structures such those that can self-consistently estimate the uncertainty of the predictions (e.g. \citealt{obrien--23}).

\par

%

\onecolumn

\section{Additional tables}
\label{sect:appdx:tables}

\begin{table*}
\centering
\caption{Best-fitting parameters for SN~2011fe and SN~2013dy with W7 models and different weighting schemes}\tabularnewline
\label{tab:masks_spec_log_w7}\tabularnewline
\resizebox{\textwidth}{!}{
\begin{tabular}{llllllllll}\hline
\hline
\tabularnewline[-0.25cm]
& \multicolumn{3}{c}{Uniform} & \multicolumn{3}{c}{Blue only} & \multicolumn{3}{c}{Features only} \tabularnewline
Phase  &  Luminosity        & Inner boundary  &  $\ln \mathcal{L}(\theta)$ & Luminosity        & Inner boundary  & $\ln \mathcal{L}(\theta|o)$ & Luminosity        & Inner boundary &  $\ln \mathcal{L}(\theta|o)$ \tabularnewline
(days) &  (log $L_{\odot}$) & (km~s$^{-1}$)   &                            & (log $L_{\odot}$) & (km~s$^{-1}$)   &                           & (log $L_{\odot}$) & (km~s$^{-1}$)  &         \tabularnewline
\hline
\multicolumn{10}{c}{SN~2011fe} \tabularnewline
\hline
\phn$-$13.1     &  ${8.56^{8.58}_{8.54} }$ & ${12221^{12598}_{11623} }$     & ${-88349_{-88357}^{-88330} }$    & ${8.57^{8.59}_{8.56} }$ & ${12909^{13315}_{12579} }$ & ${-62153_{-62156}^{-62150} }$  & ${8.54^{8.57}_{8.51} }$ & ${13992^{14996}_{12167} }$   & ${-15628_{-15630}^{-15627} }$ \vspace{0.2cm}	    \tabularnewline 
\phn$-$10.1     &  ${9.11^{9.12}_{9.09} }$ & ${12181^{12589}_{11591} }$     & ${-89153_{-89158}^{-89148} }$    & ${9.12^{9.14}_{9.11} }$ & ${12858^{13228}_{12415} }$ & ${-62833_{-62835}^{-62827} }$  & ${9.06^{9.13}_{9.03} }$ & ${12366^{13678}_{11720} }$   & ${-15734_{-15736}^{-15731} }$ \vspace{0.2cm}	   \tabularnewline 
\phn$-$6.9      &  ${9.40^{9.42}_{9.38} }$ & \phn{}${6740^{6956}_{6671} }$ 	& ${-89730_{-89735}^{-89727} }$    & ${9.43^{9.44}_{9.42} }$ & ${12210^{12886}_{11492} }$ & ${-63199_{-63201}^{-63198} }$  & ${9.40^{9.44}_{9.36} }$ & ${12115^{12757}_{10368} }$   & ${-15803_{-15804}^{-15802} }$ \vspace{0.2cm}	    \tabularnewline 
\phn$-$2.9      &  ${9.53^{9.54}_{9.51} }$ & \phn{}${5730^{6185}_{5330} }$  & ${-90022_{-90024}^{-90020} }$    & ${9.57^{9.58}_{9.56} }$ & ${11445^{12140}_{10775} }$ & ${-63426_{-63427}^{-63425} }$  & ${9.55^{9.58}_{9.52} }$ & ${10888^{11924}_{9847} }$	& ${-15857_{-15858}^{-15856} }$ \vspace{0.2cm}	    \tabularnewline 
\phn$+$0.1      &  ${9.54^{9.56}_{9.52} }$ & \phn{}${4840^{5320}_{4350} }$	& ${-90225_{-90227}^{-90222} }$    & ${9.59^{9.61}_{9.58} }$ & ${11080^{11635}_{10239} }$ & ${-63485_{-63487}^{-63480} }$  & ${9.58^{9.61}_{9.55} }$ & ${10369^{11097}_{9647} }$	& ${-15871_{-15872}^{-15871} }$ \vspace{0.2cm}	   \tabularnewline 
\phn$+$3.4      &  ${9.50^{9.53}_{9.48} }$ & \phn{}${4033^{4338}_{4000} }$	& ${-90520_{-90525}^{-90514} }$    & ${9.57^{9.59}_{9.56} }$ & ${10311^{11001}_{9666} }$  & ${-63431_{-63435}^{-63425} }$  & ${9.57^{9.60}_{9.54} }$ & ${10058^{10719}_{9178} }$	& ${-15873_{-15873}^{-15872} }$ \vspace{0.2cm}	    \tabularnewline 
\hline
               &                          & $\ln \mathcal{Z}$          & ${-538059_{-538082}^{-538032} }$ &                         & $\ln \mathcal{Z}$          & ${-378589_{-378597}^{-378578} }$ &                       & $\ln \mathcal{Z}$            & ${-94816_{-94819}^{-94814} }$ \tabularnewline
\hline\hline
\multicolumn{10}{c}{SN~2013dy} \tabularnewline
\hline
\phn$-$6.6     & ${9.42^{9.44}_{9.40} }$ & \phn{}${8956^{9490}_{8404} }$ & ${-89846_{-89853}^{-89835} }$ & ${9.45^{9.51}_{9.43} }$ & ${11340^{12577}_{10226} }$         & ${-63309_{-63315}^{-63300} }$ & ${9.52^{9.54}_{9.46} }$ & ${10778^{14242}_{10097} }$	        & ${-15755_{-15756}^{-15754} }$ \vspace{0.2cm}	    \tabularnewline 
\phn$-$2.5     & ${9.50^{9.53}_{9.48} }$ & \phn{}${5307^{5519}_{5009} }$ & ${-89943_{-89948}^{-89936} }$ & ${9.51^{9.52}_{9.49} }$ & \phn{}${5337^{5885}_{4884} }$ 	  & ${-63428_{-63433}^{-63420} }$ & ${9.55^{9.57}_{9.52} }$ & ${10569^{11466}_{9733} }$		        & ${-15811_{-15814}^{-15809} }$ \vspace{0.2cm}	    \tabularnewline 
\phn$-$0.8     & ${9.51^{9.54}_{9.48} }$ & \phn{}${4465^{4613}_{4257} }$ & ${-89945_{-89954}^{-89939} }$ & ${9.50^{9.52}_{9.49} }$ & \phn{}${5306^{5790}_{4853} }$ 	  & ${-63455_{-63457}^{-63452} }$ & ${9.54^{9.56}_{9.52} }$ & ${10438^{11255}_{9633} }$		        & ${-15807_{-15808}^{-15805} }$ \vspace{0.2cm}	    \tabularnewline 
\phn$+$1.2     & ${9.48^{9.50}_{9.46} }$ & \phn{}${4003^{4066}_{4000} }$ & ${-90012_{-90019}^{-90001} }$ & ${9.48^{9.50}_{9.47} }$ & \phn{}${5282^{5733}_{4834} }$ 	  & ${-63474_{-63476}^{-63472} }$ & ${9.52^{9.55}_{9.51} }$ & ${10188^{10858}_{9473} }$		        & ${-15803_{-15804}^{-15802} }$ \vspace{0.2cm}	    \tabularnewline 
\phn$+$4.5     & ${9.41^{9.43}_{9.39} }$ & \phn{}${4001^{4026}_{4000} }$ & ${-90241_{-90250}^{-90232} }$ & ${9.44^{9.46}_{9.43} }$ & \phn{}${5247^{5698}_{4818} }$ 	  & ${-63454_{-63459}^{-63451} }$ & ${9.48^{9.50}_{9.45} }$ & \phn{}${9912^{10506}_{9060} }$		& ${-15810_{-15811}^{-15809} }$ \vspace{0.2cm}	    \tabularnewline 
\hline
         &                         & $\ln \mathcal{Z}$       & ${-450039_{-450050}^{-450024} }$ &                         & $\ln \mathcal{Z}$       & ${-317169_{-317175}^{-317157} }$ &                       & $\ln \mathcal{Z}$            &  ${-79031_{-79034}^{-79028} }$ \tabularnewline
\hline\hline
\multicolumn{9}{p{17cm}}{\textit{Note.} The best-fitting model parameters are determined based on the median of the total posterior across all of the neural networks included in the fits. Similarly, upper and lower limits are also based on the total posterior across all neural networks. We stress that these parameters assume uniform wavelength weighting, which has important limitations (see Sect.~\ref{sect:weightings}).}  \tabularnewline
\end{tabular}
}
\end{table*}

\begin{table*}
\centering
\caption{Best-fitting parameters for SN~2011fe and SN~2013dy with N100 models and different weighting schemes}\tabularnewline
\label{tab:masks_spec_log_n100}\tabularnewline
\resizebox{\textwidth}{!}{
\begin{tabular}{llllllllll}\hline
\hline
\tabularnewline[-0.25cm]
 & \multicolumn{3}{c}{Uniform} & \multicolumn{3}{c}{Blue only} & \multicolumn{3}{c}{Features only} \tabularnewline
Phase  &  Luminosity        & Inner boundary  & $\ln \mathcal{L}(\theta|o)$ &  Luminosity        & Inner boundary  & $\ln \mathcal{L}(\theta|o)$ &  Luminosity        & Inner boundary & $\ln \mathcal{L}(\theta|o)$  \tabularnewline
(days) &  (log $L_{\odot}$) & (km~s$^{-1}$)   &                           &  (log $L_{\odot}$) & (km~s$^{-1}$)   &                           &  (log $L_{\odot}$) & (km~s$^{-1}$)  &         \tabularnewline
\hline
\multicolumn{10}{c}{SN~2011fe} \tabularnewline
\hline
\phn$-$13.1     & ${8.56^{8.58}_{8.55} }$ & ${13987^{14348}_{13583} }$  & ${-88263_{-88273}^{-88252} }$      & ${8.58^{8.60}_{8.56} }$ & ${13554^{14001}_{13093} }$       & ${-62107_{-62115}^{-62098} }$ & ${8.57^{8.61}_{8.54} }$ & ${16084^{16819}_{15095} }$ & ${-15556_{-15563}^{-15551} }$	\vspace{0.2cm}	    \tabularnewline 
\phn$-$10.1     & ${9.10^{9.11}_{9.08} }$ & ${12359^{13013}_{11828} }$  & ${-89197_{-89200}^{-89194} }$      & ${9.12^{9.14}_{9.10} }$ & ${12550^{13265}_{11902} }$       & ${-62831_{-62833}^{-62827} }$ & ${9.11^{9.14}_{9.07} }$ & ${13389^{15594}_{12463} }$ & ${-15687_{-15690}^{-15684} }$	\vspace{0.2cm}	   \tabularnewline 
\phn$-$6.9      & ${9.39^{9.41}_{9.38} }$ & ${11267^{11995}_{10829} }$  & ${-89785_{-89788}^{-89783} }$      & ${9.41^{9.43}_{9.40} }$ & ${11022^{11679}_{10541} }$       & ${-63180_{-63182}^{-63179} }$ & ${9.41^{9.43}_{9.38} }$ & ${12546^{13611}_{11634} }$ & ${-15777_{-15778}^{-15775} }$	\vspace{0.2cm}	    \tabularnewline 
\phn$-$2.9      & ${9.52^{9.54}_{9.51} }$ & \phn{}${7225^{7738}_{6837} }$  	& ${-90248_{-90253}^{-90239} }$  & ${9.55^{9.57}_{9.54} }$ & ${10456^{10821}_{10121} }$       & ${-63411_{-63415}^{-63409} }$ & ${9.55^{9.58}_{9.53} }$ & ${11602^{12453}_{10865} }$ & ${-15853_{-15855}^{-15851} }$	\vspace{0.2cm}	    \tabularnewline 
\phn$+$0.1      & ${9.52^{9.54}_{9.50} }$ & \phn{}${4035^{4340}_{4000} }$  	& ${-90288_{-90295}^{-90280} }$  & ${9.58^{9.59}_{9.57} }$ & ${10130^{10467}_{9840} }$        & ${-63504_{-63508}^{-63500} }$ & ${9.60^{9.62}_{9.58} }$ & ${10955^{11832}_{10423} }$ & ${-15905_{-15908}^{-15903} }$	\vspace{0.2cm}	   \tabularnewline 
\phn$+$3.4      & ${9.49^{9.51}_{9.47} }$ & \phn{}${4011^{4249}_{4000} }$  	& ${-90526_{-90538}^{-90517} }$  & ${9.56^{9.57}_{9.55} }$ & \phn{}${9931^{10294}_{9405} }$   & ${-63525_{-63528}^{-63520} }$ & ${9.62^{9.64}_{9.59} }$ & ${10471^{11679}_{9719} }$  & ${-15979_{-15981}^{-15975} }$	\vspace{0.2cm}	    \tabularnewline 
\hline
               &                          & $\ln \mathcal{Z}$          & ${-538366_{-538377}^{-538351} }$ &                         & $\ln \mathcal{Z}$          & ${-378622_{-378636}^{-378609} }$ &                       & $\ln \mathcal{Z}$         & ${-94811_{-94821}^{-94801} }$ \tabularnewline
\hline\hline
\multicolumn{10}{c}{SN~2013dy} \tabularnewline
\hline
\phn$-$6.6     &  ${9.41^{9.45}_{9.39} }$ & ${11167^{13222}_{10325} }$          & ${-89850_{-89868}^{-89833} }$  & ${9.45^{9.48}_{9.43} }$ & ${11987^{12989}_{11306} }$ & ${-63246_{-63251}^{-63241} }$  &  ${9.45^{9.49}_{9.41} }$ & ${14243^{16277}_{12084} }$       & ${-15765_{-15775}^{-15756} }$  \vspace{0.2cm}	    \tabularnewline 
\phn$-$2.5     &  ${9.48^{9.51}_{9.46} }$ & ${10422^{12155}_{9206} }$  	        & ${-90287_{-90330}^{-90254} }$  & ${9.52^{9.54}_{9.51} }$ & ${10888^{11306}_{10563} }$ & ${-63438_{-63441}^{-63437} }$  &  ${9.53^{9.56}_{9.51} }$ & ${12379^{13534}_{11808} }$       & ${-15809_{-15810}^{-15808} }$  \vspace{0.2cm}	    \tabularnewline 
\phn$-$0.8     &  ${9.48^{9.51}_{9.47} }$ & \phn{}${9961^{11743}_{8510} }$  	& ${-90451_{-90474}^{-90432} }$  & ${9.53^{9.54}_{9.51} }$ & ${10622^{10983}_{10258} }$ & ${-63462_{-63466}^{-63460} }$  &  ${9.54^{9.56}_{9.51} }$ & ${12141^{12866}_{11382} }$       & ${-15830_{-15835}^{-15826} }$  \vspace{0.2cm}	    \tabularnewline 
\phn$+$1.2     &  ${9.47^{9.50}_{9.45} }$ & \phn{}${9293^{11346}_{7572} }$  	& ${-90559_{-90569}^{-90530} }$  & ${9.52^{9.53}_{9.51} }$ & ${10309^{10712}_{9930} }$  & ${-63478_{-63482}^{-63474} }$  &  ${9.54^{9.56}_{9.51} }$ & ${11846^{12808}_{10798} }$       & ${-15875_{-15883}^{-15869} }$ \vspace{0.2cm}	    \tabularnewline 
\phn$+$4.5     &  ${9.42^{9.44}_{9.40} }$ & \phn{}${5363^{6417}_{4000} }$  	    & ${-90564_{-90596}^{-90462} }$  & ${9.49^{9.51}_{9.48} }$ & ${10075^{10566}_{9585} }$  & ${-63467_{-63471}^{-63463} }$  &  ${9.52^{9.56}_{9.48} }$ & \phn{}${8005^{12788}_{4000} }$   & ${-16020_{-16073}^{-15987} }$  \vspace{0.2cm}	    \tabularnewline 
\hline
\hline
               &                          & $\ln \mathcal{Z}$          & ${-451751_{-451762}^{-451739} }$ &                         & $\ln \mathcal{Z}$          & ${-317145_{-317148}^{-317141} }$ &                       & $\ln \mathcal{Z}$          & ${-79316_{-79322}^{-79312} }$ \tabularnewline
\hline\hline
\multicolumn{9}{p{17cm}}{\textit{Note.} The best-fitting model parameters are determined based on the median of the total posterior across all of the neural networks included in the fits. Similarly, upper and lower limits are also based on the total posterior across all neural networks. We stress that these parameters assume uniform wavelength weighting, which has important limitations (see Sect.~\ref{sect:weightings}).}   \tabularnewline
\end{tabular}
}
\end{table*}

\section{Additional figures for SN~2011fe}
\label{sect:appdx:figures_11fe}

\begin{figure*}
\centering
\includegraphics[width=\textwidth]{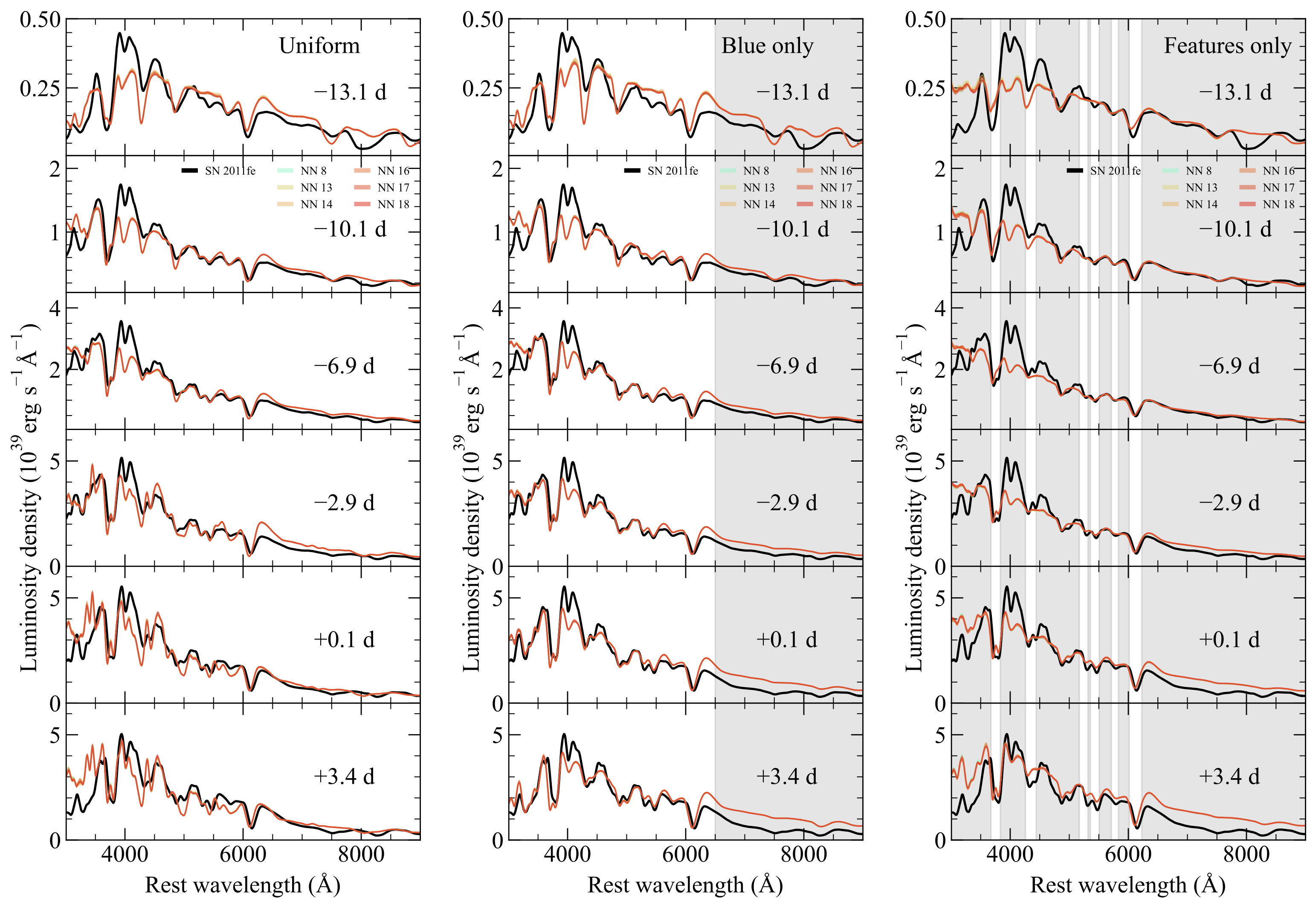}
\caption{As in Fig.~\ref{fig:11fe_w7_all_mask} for SN~2011fe and N100 models.}
\label{fig:11fe_n100_all_mask}
\centering
\end{figure*}

\begin{figure*}
\centering
\includegraphics[width=\textwidth]{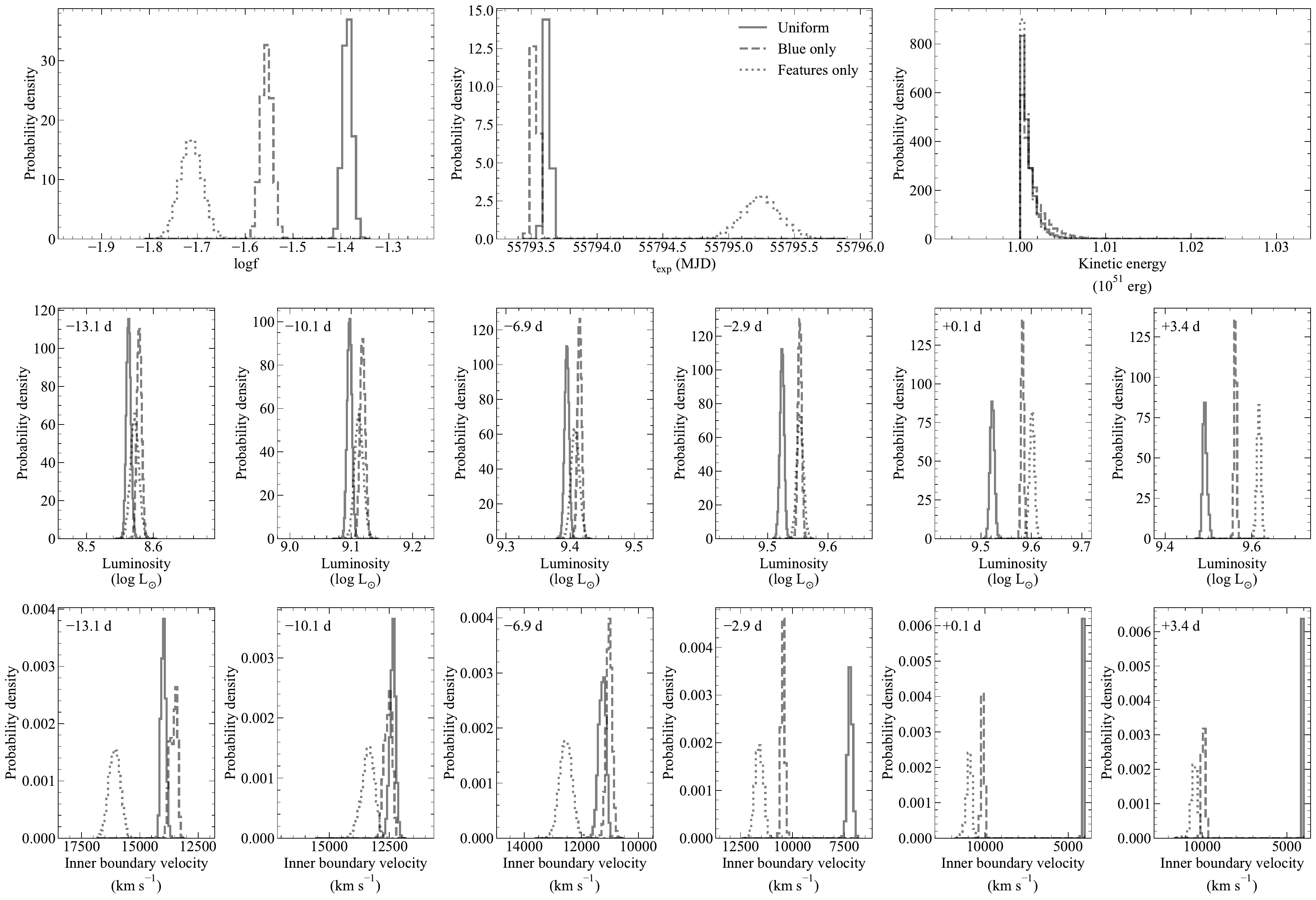}
\caption{As in Fig.~\ref{fig:11fe_w7_all_mask_posterior} for SN~2011fe and N100 models.}
\label{fig:11fe_n100_all_mask_posterior}
\centering
\end{figure*}

\section{Additional figures for SN~2013dy}
\label{sect:appdx:figures_13dy}

\begin{figure}
\centering
\includegraphics[width=\textwidth]{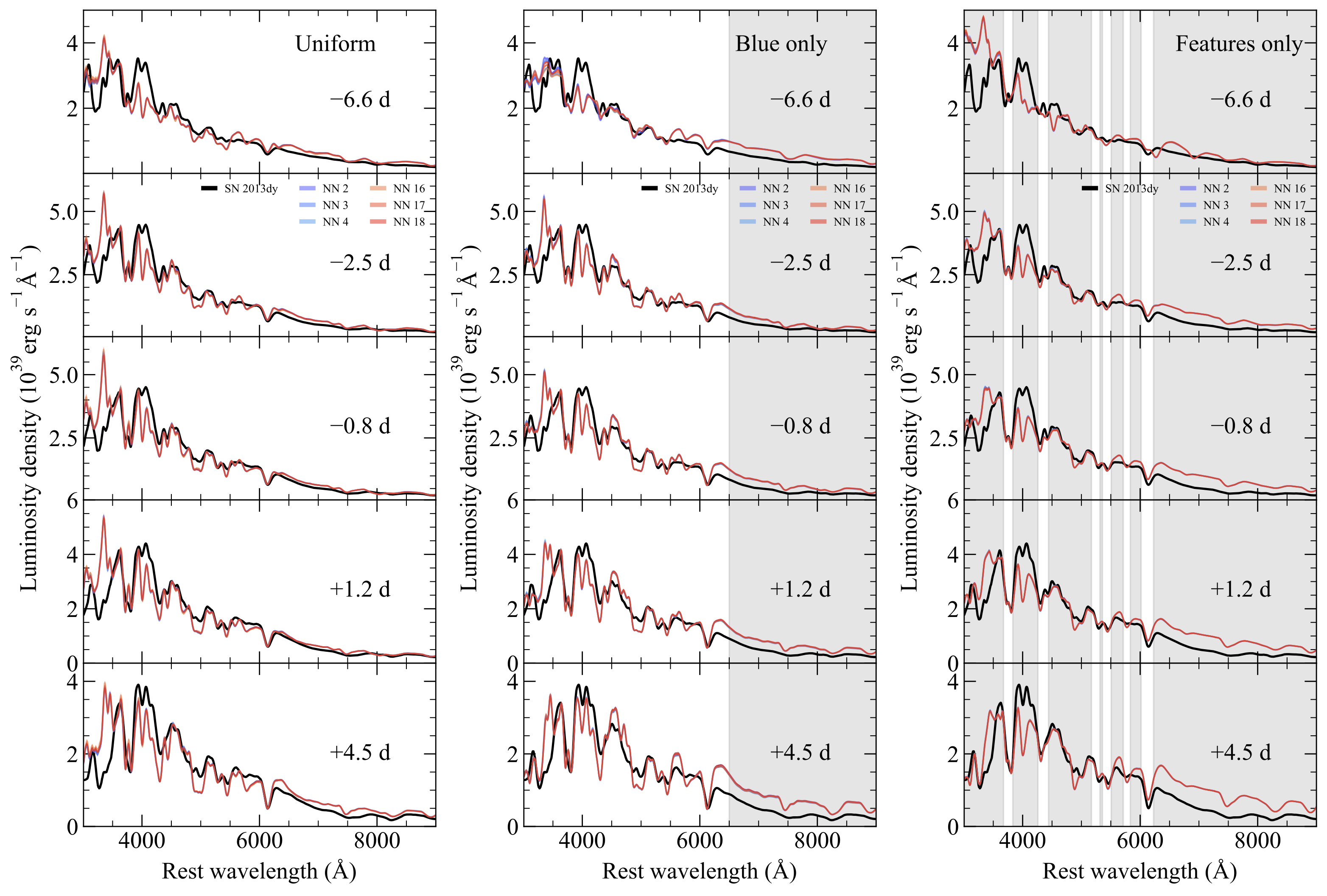}
\caption{Comparison between SN~2013dy and neural network reconstructions of the best-fitting W7 model spectra for each neural network and assuming different weighting schemes. Spectra are shown on an absolute luminosity scale with no additional scaling or offsets applied. Phases are given relative to $B$-band maximum. Grey shaded regions are not included during fitting.}
\label{fig:13dy_w7_all_mask}
\centering
\end{figure}

\begin{figure}
\centering
\includegraphics[width=\textwidth]{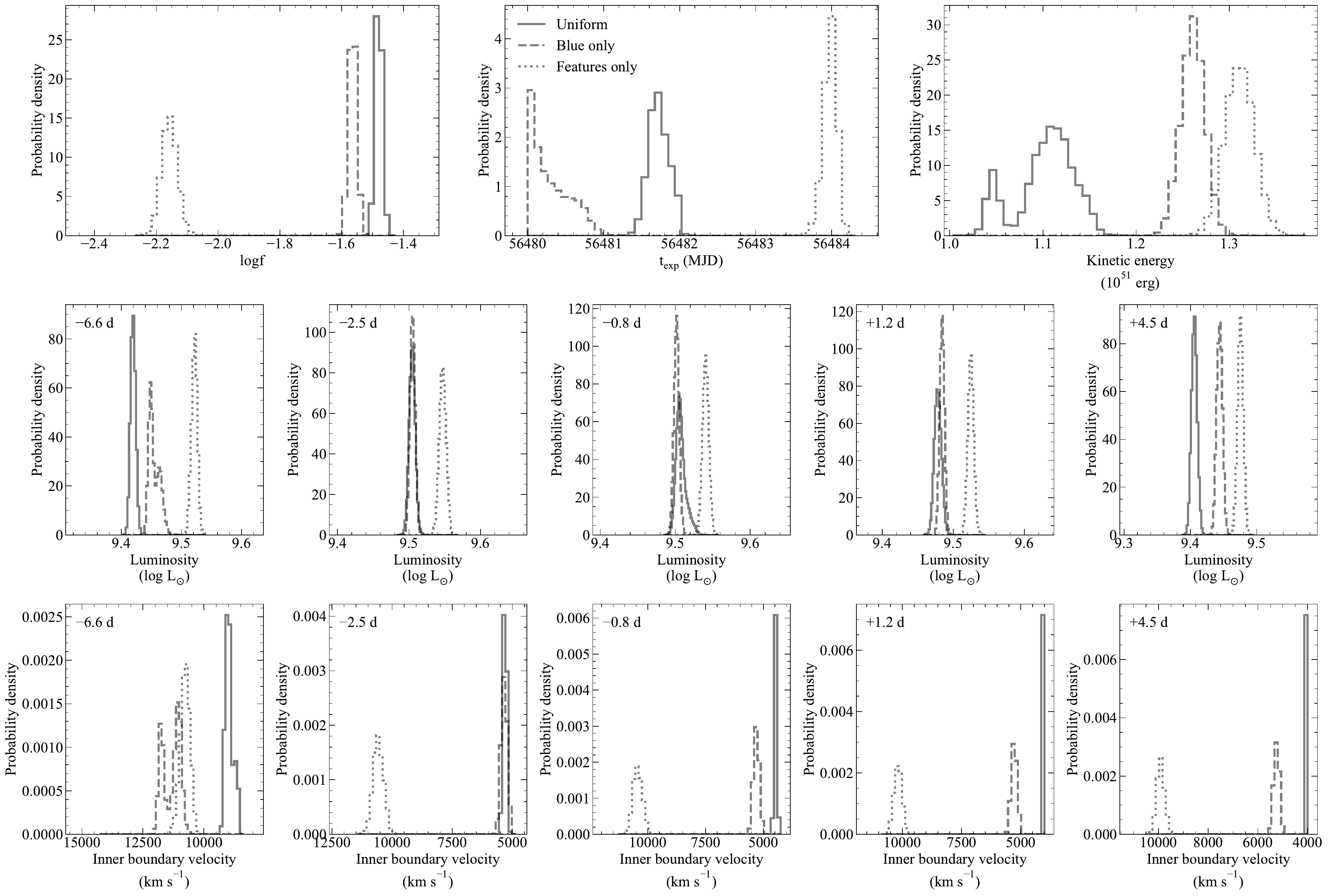}
\caption{Total posterior distributions for best-fitting parameters to SN~2013dy using W7 models across all neural networks fit and assuming different weighting schemes. Phases are given relative to $B$-band maximum.}
\label{fig:13dy_w7_all_mask_posterior}
\centering
\end{figure}

\begin{figure}
\centering
\includegraphics[width=\textwidth]{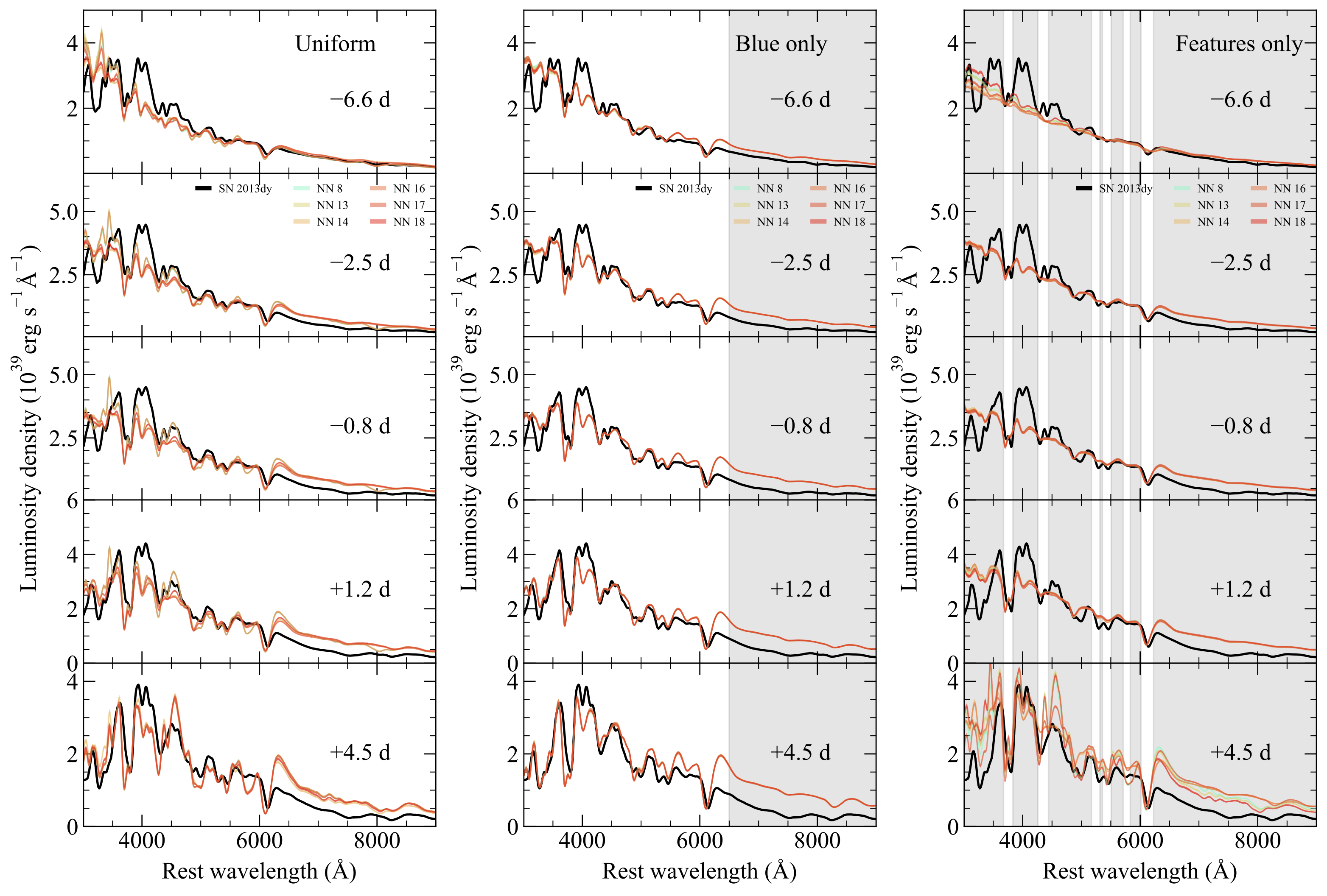}
\caption{As in Fig.~\ref{fig:13dy_w7_all_mask} for SN~2013dy and N100 models.}
\label{fig:13dy_n100_all_mask}
\centering
\end{figure}

\begin{figure}
\centering
\includegraphics[width=\textwidth]{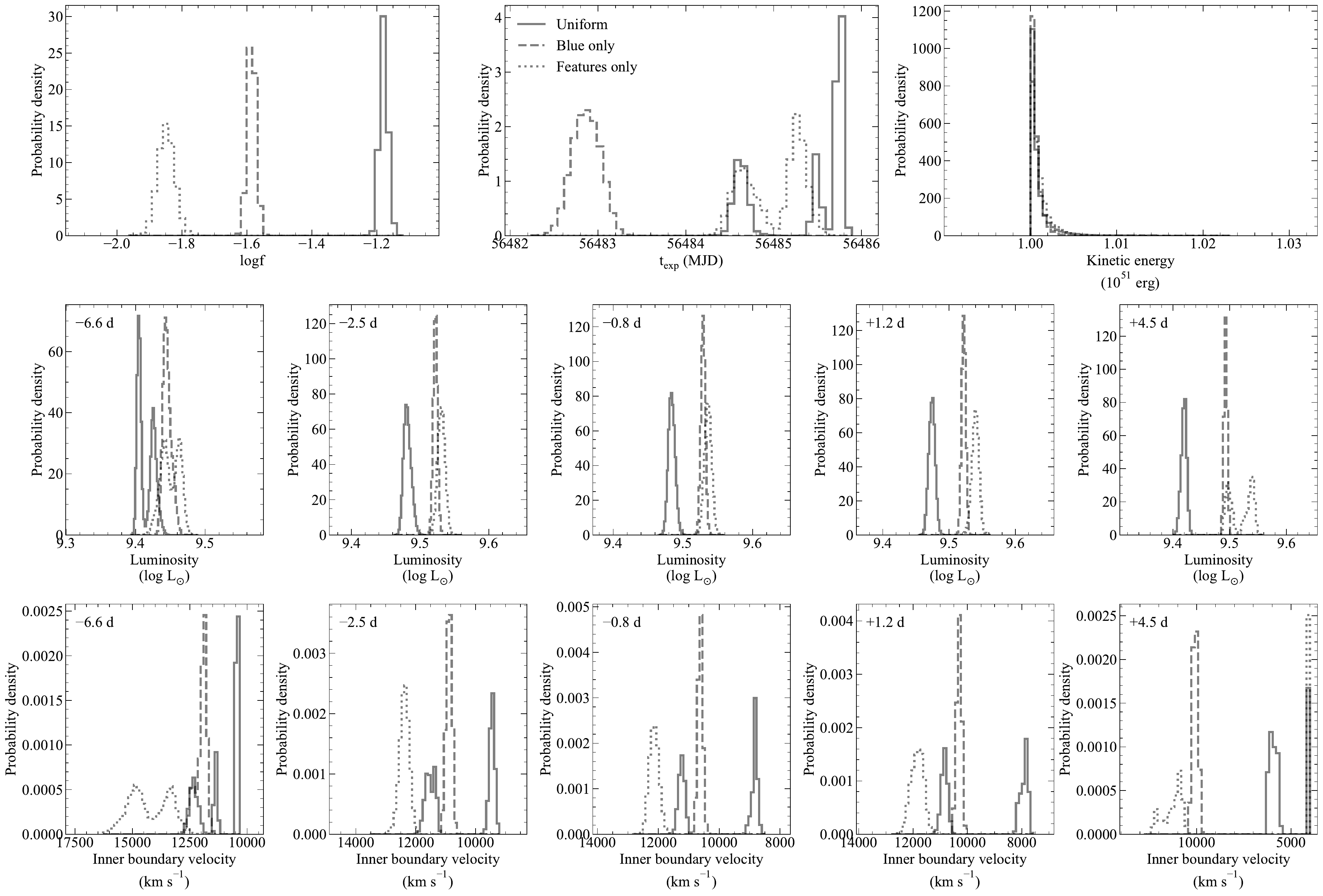}
\caption{As in Fig.~\ref{fig:13dy_w7_all_mask_posterior} for SN~2013dy and N100 models.}
\label{fig:13dy_n100_all_mask_posterior}
\centering
\end{figure}


\bsp	
\label{lastpage}
\end{document}